\documentclass[10pt,letterpaper]{article}
\usepackage[margin=3cm]{geometry}

\usepackage{amsmath,amssymb}

\usepackage{changepage}

\usepackage[utf8x]{inputenc}

\usepackage{textcomp,marvosym}

\usepackage{cite}

\usepackage{nameref,hyperref}

\usepackage[right]{lineno}

\usepackage{microtype}
\DisableLigatures[f]{encoding = *, family = * }

\usepackage[table]{xcolor}

\usepackage{array}

\newcolumntype{+}{!{\vrule width 2pt}}

\newlength\savedwidth

\usepackage[aboveskip=1pt,labelfont=bf,labelsep=period,justification=raggedright,singlelinecheck=off]{caption}

\makeatletter
\renewcommand{\@biblabel}[1]{\quad#1.}
\makeatother

\usepackage{lastpage,fancyhdr,graphicx}
\usepackage{epstopdf}
\fancyheadoffset[L]{2.25in}
\fancyfootoffset[L]{2.25in}
  \graphicspath{{./Figs4/}}

\usepackage{epstopdf}

\begin{document}
\vspace*{0.2in}

\begin{flushleft}
{\Large
\textbf\newline{Phylogenetic correlations can suffice to infer protein partners from sequences} 
}
\newline
\\
Guillaume Marmier\textsuperscript{1,2},
Martin Weigt\textsuperscript{2},
Anne-Florence Bitbol\textsuperscript{1*}
\\
\bigskip
\textbf{1} Sorbonne Universit{\'e}, CNRS, Laboratoire Jean Perrin (UMR 8237), F-75005 Paris, France
\\
\textbf{2} Sorbonne Universit{\'e}, CNRS, Institut de Biologie Paris-Seine, Laboratoire de Biologie Computationnelle et Quantitative (LCQB, UMR 7238), F-75005 Paris, France
\\
\bigskip

* anne-florence.bitbol@sorbonne-universite.fr

\end{flushleft}
\section*{Abstract}
Determining which proteins interact together is crucial to a systems-level understanding of the cell. Recently, algorithms based on Direct Coupling Analysis (DCA) pairwise maximum-entropy models have allowed to identify interaction partners among paralogous proteins from sequence data. This success of DCA at predicting protein-protein interactions could be mainly based on its known ability to identify pairs of residues that are in contact in the three-dimensional structure of protein complexes and that coevolve to remain physicochemically complementary. However, interacting proteins possess similar evolutionary histories. What is the role of purely phylogenetic correlations in the performance of DCA-based methods to infer interaction partners? To address this question, we employ controlled synthetic data that only involve phylogeny and no interactions or contacts. We find that DCA accurately identifies the pairs of synthetic sequences that share evolutionary history. While phylogenetic correlations confound the identification of contacting residues by DCA, they are thus useful to predict interacting partners among paralogs. We find that DCA performs as well as phylogenetic methods to this end, and slightly better than them with large and accurate training sets. Employing DCA or phylogenetic methods within an Iterative Pairing Algorithm (IPA) allows to predict pairs of evolutionary partners without a training set. We further demonstrate the ability of these various methods to correctly predict pairings among real paralogous proteins with genome proximity but no known direct physical interaction, illustrating the importance of phylogenetic correlations in natural data. However, for physically interacting and strongly coevolving proteins, DCA and mutual information outperform phylogenetic methods. We finally discuss how to distinguish physically interacting proteins from proteins that only share a common evolutionary history.

\section*{Author summary}
Many biologically important protein-protein interactions are conserved over evolutionary time scales. This leads to two different signals that can be used to computationally predict interactions between protein families and to identify specific interaction partners. First, the shared evolutionary history leads to highly similar phylogenetic relationships between interacting proteins of the two families. Second, the need to keep the interaction surfaces of partner proteins biophysically compatible causes a correlated amino-acid usage of interface residues. Employing simulated data, we show that the shared history alone can be used to detect partner proteins. Similar accuracies are achieved by algorithms comparing phylogenetic relationships and by methods based on Direct Coupling Analysis (DCA), which are primarily known for their ability to detect the second type of signal. Using natural sequence data, we show that in cases with shared evolutionary history but without known physical interactions, both methods work with similar accuracy, while for some physically interacting systems, DCA and mutual information outperform phylogenetic methods. We propose methods allowing both to predict interactions between protein families and to find interacting partners among paralogs.


\section*{Introduction}

The vast majority of cellular processes are carried out by interacting proteins. Functional interactions between proteins allow multi-protein complexes to properly assemble, and ensure the specificity of signal transduction pathways. Hence, mapping functional protein-protein interactions is a crucial question in biology. Since high-throughput experiments remain challenging~\cite{Rajagopala14}, an attractive alternative is to exploit the growing amount of sequence data in order to identify functional protein-protein interaction partners.

The amino-acid sequences of interacting proteins are correlated, both because of evolutionary constraints arising from the need to maintain physico-chemical complementarity among contacting amino-acids, and because of shared evolutionary history. The first type of correlations has recently received substantial interest, both within single proteins and across protein partners. Global statistical models built from the observed sequence correlations using the maximum entropy principle~\cite{Jaynes57,Lapedes99,Burger08,Weigt09}, and assuming pairwise interactions, known as Direct Coupling Analysis (DCA), have been used with success to determine three-dimensional protein structures from sequences~\cite{Marks11,Morcos11,Sulkowska12}, to analyze mutational effects~\cite{Dwyer13,Cheng14,Cheng16,Figliuzzi16} and conformational changes~\cite{Morcos13,Malinverni15}, to find residue contacts between known interaction partners~\cite{Weigt09,Procaccini11,Baldassi14,Ovchinnikov14,Hopf14,Tamir14,dosSantos15,Feinauer16}, and most recently to predict interaction partners among paralogs from sequence data~\cite{Bitbol16,Gueudre16}. Similar global statistical models have also revealed functional relationships in other contexts~\cite{Lezon06,Jiang17}. The success of DCA-based approaches at predicting protein-protein interactions~\cite{Bitbol16,Gueudre16} could originate only from correlations between residues that are in direct contact in the three-dimensional protein complex structure, thus needing to maintain physico-chemical complementarity. However, additional correlations arise in protein sequences due to their common evolutionary history, i.e. phylogeny~\cite{Casari95,Halabi09,Qin18}, even in the absence of structural constraints. Functionally related protein families~\cite{Fryxell96}, especially interacting ones~\cite{Goh00} tend to have similar phylogenies, and methods directly based on phylogeny and on sequence similarity~\cite{Pazos01,Jothi05,Bradde10,Ochoa10,Ochoa15}, in particular the Mirrortree method~\cite{Pazos01,Ochoa10,Ochoa15} allow to predict which protein families interact. The similarities in the phylogenies of interacting protein families can arise from the coevolution of residues in structural contact, but also from more global shared evolutionary pressures, resulting in similar evolutionary rates~\cite{Hakes07,Juan08,Kann09,Lovell10,Swapna12}, and from shared evolutionary history unrelated to constraints, including common timing of speciation and gene duplication events~\cite{Lovell10}. While being detrimental to the identification of contacting residues by DCA~\cite{Weigt09,Marks11,Qin18}, these additional sources of signal can aid the identification of interaction partners. Accordingly, a method based on mutual information (MI) was recently shown to slightly outperform the DCA-based one~\cite{BitbolPMI}. MI includes all types of statistical dependence between the sequences of interacting partners.

To what extent do purely phylogenetic correlations contribute to the prediction of interaction partners from sequences by DCA? If sequences only share a common evolutionary history, i.e. in the absence of functional constraints, how do DCA-based methods compare to phylogenetic methods? Answering these questions is important to understand the reasons of the success of DCA-based methods, and will open the way to developing new methods that combine useful information from both phylogeny and contacts. To address these questions, we generate controlled synthetic data that only involve phylogeny, in the absence of functional constraints. Our DCA-based method correctly identifies pairs of synthetic ``sequences'' that share evolutionary history, even without any training set, thanks to an Iterative Pairing Algorithm (IPA). (Strikingly, this high predictive power is obtained in the absence of real couplings from interactions, from purely phylogenetic correlations.) On this synthetic dataset, we find that the DCA-based IPA and a phylogeny-based IPA reach similar performances, with DCA slightly outperforming the phylogenetic method for large training sets. We then show examples of natural proteins without known direct physical interactions but with shared evolutionary history that can be accurately paired by our various methods, thus illustrating the importance of phylogenetic correlations in real data. For a pair of actually interacting and strongly coevolving protein families, we find that DCA and MI substantially outperform phylogenetic methods. Finally, we propose methods to predict protein-protein interactions from the level of protein families to that of paralogs. 

\section*{Methods}

\subsection*{Synthetic data generation}

\begin{figure}[h!]
	\centering
	\includegraphics[width=0.9\textwidth]{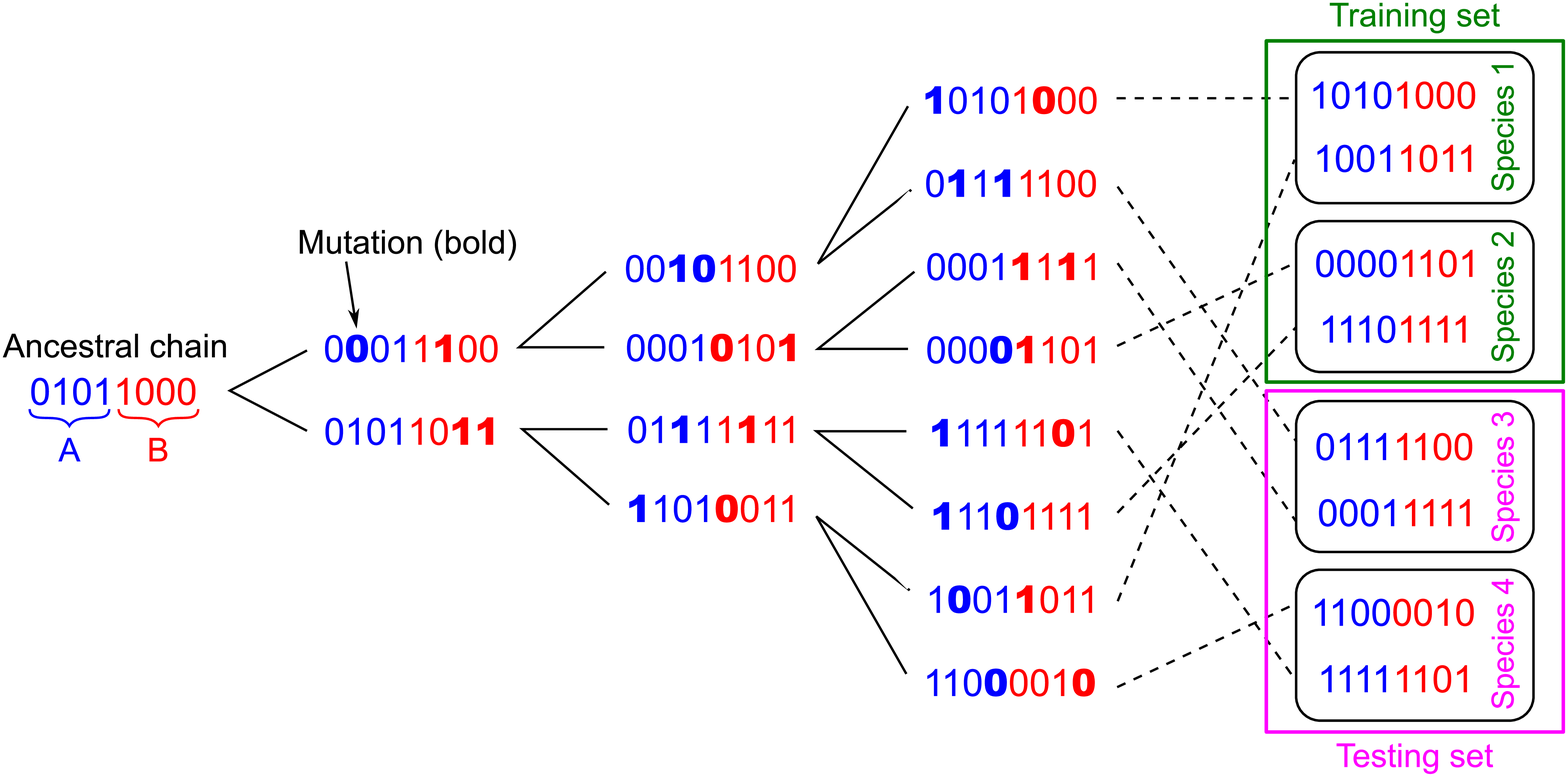}
	\vspace{0.2cm}
	\caption{{\bf Construction of a synthetic dataset of chains sharing evolutionary history.} Starting from a random ancestral chain AB of bits whose two halves A and B are shaded in blue and red, a series of $n$ duplication and mutation steps (``generations'', here $n=3$) are performed (bold: mutated bits; here 2 bits per chain are mutated at each step), resulting in $2^n=8$ chains. Species are then constructed randomly, here with $m=2$ chains per species. Some species are considered as the training set (green), and the other ones constitute the testing set (pink), where the pairings between each chain A and each chain B will be blinded. }
	\label{Fig1}
\end{figure}

We generate controlled synthetic data where ``sequences'' are modeled as strings of binary variables (bits) taking values 0 or 1. In real protein sequences, each site can feature 21 states (20 amino acids, plus the alignment gap), but binary models where the consensus or reference amino acid is denoted by 0 and mutant states by 1 retain all conceptual ingredients, and have proved useful to identify sectors of collectively correlated amino acids~\cite{Halabi09,Dahirel11}, as well as to predict fitness landscapes from sets of closely related proteins~\cite{Mann14}. Our synthetic sequences are evolved along a phylogenetic tree represented by a branching process with random mutations, in the absence of any constraint stemming from interactions or function. Hence, all correlations in this synthetic data arise from shared evolutionary history (and finite-size noise). The data generation process is illustrated in Fig.~\ref{Fig1}.

Specifically, we consider perfect binary trees for simplicity. The ancestral chain, composed of uniformly randomly distributed bits, is duplicated, giving rise to two chains, and mutations are performed independently in the two duplicate chains: each mutation changes the state of one uniformly randomly chosen bit of the chain. Then, the new chains are duplicated again, and so on. We employ two different models for the occurrence of mutations. In the simplest model, a fixed number of mutations per total chain length is performed along each branch of the tree, i.e. between two duplication steps. In a more realistic model, the number of mutations per branch is drawn in a Poisson distribution with fixed mean. After a given number $n$ of duplication steps (``generations'', representing ancestry in terms of speciation or gene duplication events), a final dataset of $2^n$ chains is obtained (see Fig.~\ref{Fig1}). In practice, we take $n\leq 12$ to have tractable datasets of a few thousands of final chains. Throughout, we perform inference using these final chains, which correspond to contemporary sequences in a natural dataset. In each of these chains, the state of each bit is uniformly distributed. However, correlations exist between chains due to their shared evolutionary history. The strength of these correlations depends on the number of mutations per branch and on how close the chains are along the phylogenetic tree.

In order to introduce the notion of species in a minimal way, we randomly group chains into sets of equal size $m$, each representing a species. The $m$ different chains within a species can then be thought of as paralogs, i.e. homologs sharing common ancestry and present in the same genome. In reality, different correlations are expected between the paralogs present in a given species and the orthologs present across species. Later on, we therefore also consider another type of phylogeny that accounts for these effects, and assess the robustness of our conclusions to this variant. Note that the present minimal model with random species is realistic in the case where exchange between species (i.e. horizontal gene transfer) is sufficiently frequent.

We finally cut each chain of the final dataset in two halves of equal length. These halves, denoted by chain A and chain B, thus represent a pair of proteins that possess the same evolutionary history. Next, we blind the pairings for the chains A and B from some species (testing set) and ask whether DCA-, MI- and similarity-based methods are able to pair each A chain with its ``evolutionary partner'', namely with the B chain that possesses the same evolutionary history, starting from the known pairs (training set).

\subsection*{Inference methods}

We test several inference methods to predict pairings between chains A and B in our synthetic datasets. For each of them, performance is assessed both with a training set and without a training set. In the first case, the parameters defining scores are computed using the training set and employed to pair data in the testing set~\cite{Burger08,Procaccini11}. In the second case, we employ the Iterative Pairing Algorithm (IPA) developed in Refs.~\cite{Bitbol16,BitbolPMI} to bootstrap the predictions starting from initial random within-species pairings. Below, we present the various inference methods assuming that there is a training set. The extension to the training set-free case is then performed exactly as described in Refs.~\cite{Bitbol16,BitbolPMI}. Matlab implementations of the MI-IPA, the DCA-IPA and the Mirrortree-IPA on our standard HK-RR dataset are freely available at \url{https://doi.org/10.5281/zenodo.1421781}, \url{https://doi.org/10.5281/zenodo.1421861} and \url{https://doi.org/10.5281/zenodo.3377592} respectively.

\paragraph{Training set statistics.} To describe the statistics of a training set of synthetic paired chains AB, of total length $2L$ (where $L$ is the length of a chain A or B), we employ the empirical one-site frequencies of each state $\sigma_i\in\{0,1\}$ at each site $i\in\{1,\dots,2L\}$, denoted by $f_i(\sigma_i)$, and the two-site frequencies of occurrence of each ordered pair of states $(\sigma_i,\sigma_j)$ at each ordered pair of sites $(i,j)$, denoted by $f_{ij}(\sigma_i,\sigma_j)$. Correlations are then computed as $C_{ij}(\sigma_i,\sigma_j)=f_{ij}(\sigma_i,\sigma_j)-f_i(\sigma_i)f_j(\sigma_j)$.

\paragraph{Pseudocount.} When dealing with real protein sequences, pseudocounts are often introduced to avoid mathematical issues such as divergences due to amino-acid pairs that never appear, both with DCA~\cite{Weigt09,Procaccini11,Marks11,Morcos11} and with MI~\cite{BitbolPMI}. Introducing a pseudocount weight $\Lambda$, which effectively corresponds to adding a fraction $\Lambda$ of chains with uniformly distributed states, the corrected one-body frequencies read $\tilde{f}_i(\sigma_i)=\Lambda/2+(1-\Lambda)f_i(\sigma_i)$. Similarly, the corrected two-body frequencies read
$\tilde{f}_{ij}(\sigma_i,\sigma_j)=\Lambda/4+(1-\Lambda)f_{ij}(\sigma_i,\sigma_j)\textrm{ if }i\neq j$ and
$\tilde{f}_{ii}(\sigma_i,\sigma_j)=\delta_{\sigma_i\sigma_j}\Lambda/2 +(1-\Lambda)f_{ii}(\sigma_i,\sigma_j)= \delta_{\sigma_i\sigma_j}\tilde{f}_i(\sigma_i)$,
where $\delta_{\sigma_i\sigma_j}=1$ if $\sigma_i=\sigma_j$ and 0 otherwise. We investigated the impact of varying $\Lambda$ on the performance of pairing prediction on our synthetic data using DCA and MI. Fig.~\ref{FigS1} shows that small nonzero values of $\Lambda$ perform best for MI while larger ones improve DCA performance. Therefore, in what follows, we always took $\Lambda=0.015$ for MI and $\Lambda=0.5$ for DCA, which is the typical value used when applying DCA to real proteins~\cite{Morcos11,Marks11}.

\paragraph{DCA-based method.} 
In DCA~\cite{Weigt09,Morcos11,Marks11,Cocco18}, one starts from the empirical covariances $C_{ij}(\sigma_i,\sigma_j)$ between all pairs of sites $(i,j)$, computed on the training set. Importantly, here, we are considering paired chains AB, and $i$ and $j$ range from 1 to the total length $2L$ of such a chain. DCA is based on building a global statistical model from these covariances (and the one-body frequencies)~\cite{Weigt09,Morcos11,Marks11,Cocco18}, through the maximum entropy principle~\cite{Jaynes57}. This results in a $2L$-body probability distribution $P$ of observing a given sequence $(\sigma_1,\dots,\sigma_{2L})$ that reads $P(\sigma_1,\dots,\sigma_{2L})=\exp\left[\sum_{i<j}e_{ij}(\sigma_i,\sigma_j)+\sum_{i=1}^{2L} h_i(\sigma_i)\right]/Z$, where $Z$ is a normalization constant: this corresponds to the Boltzmann distribution associated to a Potts model with couplings $e_{ij}(\sigma_i,\sigma_j)$ and fields $h_i(\sigma_i)$~\cite{Cocco18}. Inferring the couplings and the fields that appropriately reproduce the empirical covariances is a difficult problem, known as an inverse statistical physics problem~\cite{Nguyen17}. Note that these parameters are not all independent due to the gauge degree of freedom, so one can set e.g. $h_i(0)=0$ and $e_{ij}(0,\sigma_j)=e_{ij}(\sigma_i,0)=0$ for all $i,j$ and $\sigma_i,\sigma_j$, thus leaving only $h_i(1)$ and $e_{ij}(1,1)$ to determine. Within the mean-field approximation, which will be employed throughout, these coupling strengths can be approximated by $e_{ij}(1,1)=-C^{-1}_{ij}(1,1)$~\cite{Plefka82,Morcos11,Marks11}. We then transform to the zero-sum (or Ising) gauge, yielding $e_{ij}(0,1)=e_{ij}(1,0)=-e_{ij}(0,0)=-e_{ij}(1,1)=C^{-1}_{ij}(1,1)/4$. The interest of this gauge is that it attributes the smallest possible fraction of the energy to the couplings, and the largest possible fraction to the fields~\cite{Weigt09,Ekeberg13}. Note that a fully equivalent approach is to consider sequences of Ising spins instead of bits, and to employ an Ising model. Here, we have chosen the Potts model formalism for consistency with protein sequence analysis by DCA. 

The effective interaction energy $E_{AB}$ of each possible pair AB in the testing set, constructed by concatenating a chain A and a chain B, can then be assessed via 
\begin{equation}
E_{AB}=-\sum_{i=1}^{L}\sum_{j=L+1}^{2L} e_{ij}(\sigma_i^A,\sigma_j^B)\,.
\label{energy}
\end{equation}
In real proteins, approximately minimizing such a score has proved successful at predicting interacting partners~\cite{Procaccini11,Bitbol16}. Note that we only sum over inter-chain pairs (i.e. pairs of sites involving one site in A and one in B) because we are interested in interactions between A and B. 

Importantly, DCA was designed to infer actual interactions between contacting amino acids through the couplings $e_{ij}$~\cite{Weigt09,Morcos11,Marks11,Cocco18}. By contrast, in the present synthetic data, there are no such interactions, and all correlations have their origin in phylogeny (or finite-size noise). Nevertheless, the DCA-based interaction energy in Eq.~\ref{energy} contains information about these correlations, and we will investigate how well it captures them.

\paragraph{MI-based method.} Our method based on Mutual Information (MI) was introduced in~\cite{BitbolPMI}. As with DCA, we start by describing the statistics of the training set, which is composed of complete chains AB. For this, we employ the single-site frequencies $\tilde f_i(\sigma_i)$ and the two-site frequencies $\tilde f_{ij}(\sigma_i,\sigma_j)$ (see above). The pointwise mutual information (PMI) of a pair of states $(\sigma_i,\sigma_j)$ at a pair of sites $(i,j)$ is defined as~\cite{Fano61,Church90,Role11}:
\begin{equation}
\textrm{PMI}_{ij}(\sigma_i,\sigma_j)=\log\left[\frac{\tilde f_{ij}(\sigma_i,\sigma_j)}{\tilde f_i(\sigma_i)\tilde f_j(\sigma_j)}\right]\,.
\label{PMI}
\end{equation}
Averaging this quantity over all possible pairs of states yields an estimate of the mutual information (MI) between sites $i$ and $j$~\cite{Cover06}:
$\textrm{MI}_{ij}=\sum_{\sigma_i,\sigma_j}\tilde f_{ij}(\sigma_i,\sigma_j)\,\textrm{PMI}_{ij}(\sigma_i,\sigma_j)$. 

Next, we define a pairing score $S_\mathrm{AB}$ for each possible pair AB of chains from the testing set as the sum of the PMIs of the inter-chain pairs of sites of this concatenated chain AB (i.e. those that involve one site in chain A and one site in chain B):
\begin{equation}
S_\mathrm{AB}=\sum_{i=1}^{L}\sum_{j=L+1}^{2L}\textrm{PMI}_{ij}(\sigma_i^A,\sigma_j^B)\,.
\label{SAB}
\end{equation}
In real proteins, approximately maximizing such a score has proved successful at predicting interacting partners, slightly outperforming DCA~\cite{BitbolPMI}.

\paragraph{Mirrortree-based method.} Methods based only on phylogeny and sequence similarity have been developed to predict protein-protein interactions. In particular, the Mirrortree method quantifies the similarities of distance matrices between the proteins of two families to determine whether they interact~\cite{Pazos01,Ochoa10,Ochoa15}, and has allowed the successful prediction of protein-protein interactions. This method generally relies on finding one ortholog of the proteins of interest in each species and does not address the question of which paralog of family A interacts with which paralog of family B. However, related approaches have tackled this problem~\cite{Ramani03,Gertz03,Izarzugaza06,Izarzugaza08,Bradde10,ElKebir13}, which was subsequently directly addressed by the DCA- and MI-based methods of Refs.~\cite{Bitbol16,Gueudre16,BitbolPMI}.

We introduce an approach close to the original Mirrortree algorithm~\cite{Pazos01,Ochoa10,Ochoa15} that addresses the paralog pairing problem. Specifically, let $\{A_1B_1,\dots, A_MB_M\}$ be the training set, which contains $M$ known pairs of chains. For each chain A of the testing set, we compute the vector $d_A=(d(A,A_1),\dots, d(A,A_M))$ of Hamming distances between A and each chain A of the training set. We also compute an analogous vector $d_B$ for each chain B of the training set. Next, we define a pairing score $M_{AB}$ for each possible pair AB of chains A and B from the testing set as the Pearson correlation $\rho$ between $d_A$ and $d_B$:
\begin{equation}
M_{AB}=\rho(d_A,\,d_B)\,.
\label{MAB}
\end{equation}
This score thus assigns high values to pairs AB that have highly similar phylogenetic relationships to the training set, hinting towards substantial shared evolutionary history between A and B. It can be used for predicting partnerships exactly as the DCA- and MI-based scores in Eqs.~\ref{energy} and~\ref{SAB}. Note that one then aims to maximize $M_{AB}$, just like $S_{AB}$, while DCA effective interaction energies $E_{AB}$ should be minimized.

\paragraph{Other methods based on sequence similarity.} Because there are many ways to exploit sequence similarity in order to assess shared evolutionary history, we also consider variants beyond our Mirrortree-based method. Specifically, we present results obtained using orthology between pairs, defined as reciprocal best hits in terms of Hamming distances, as well as results obtained by simply employing the Hamming distance of each possible AB pair of the testing set to its closest AB pair in the training set as a pairing score. These methods are detailed and studied in Fig.~\ref{FigS4}.

\paragraph{Pairing prediction.} We employed two different approaches to predict pairings from each of the three scores defined in Eqs.~\ref{energy},~\ref{SAB} and~\ref{MAB}. In the first approach, for each chain A, we simply picked the chain B within the same species that optimizes the pairing score. Note that this simple method is asymmetric and allows multiple chains A to be matched with the same chain B. In the second approach, we used the Hungarian algorithm (also known as the Munkres algorithm)~\cite{Kuhn55,Munkres57,HungAlg} to find the one-to-one association of each chain A with a chain B that optimizes the sum of the pairing scores within each species.

\section*{Results}

\subsection*{DCA accurately identifies pairs of chains that only share a common evolutionary history}

First, we set out to assess whether DCA can identify pairs of chains AB that only share a common evolutionary history. For this, we generated chains of bits employing a branching process with random mutations, in the absence of any interaction or functional constraint (see Methods and Fig.~\ref{Fig1}). The only correlations present among these chains thus arise from shared evolutionary history (and finite-size noise). We first ask whether DCA pairing scores (Eq.~\ref{energy}) learned on a training set of complete chains allow to correctly predict pairs of evolutionary partners in a testing set of chains separated into half chains A and B. 

\begin{figure}[h!]
	\centering
	\includegraphics[width=0.8\textwidth]{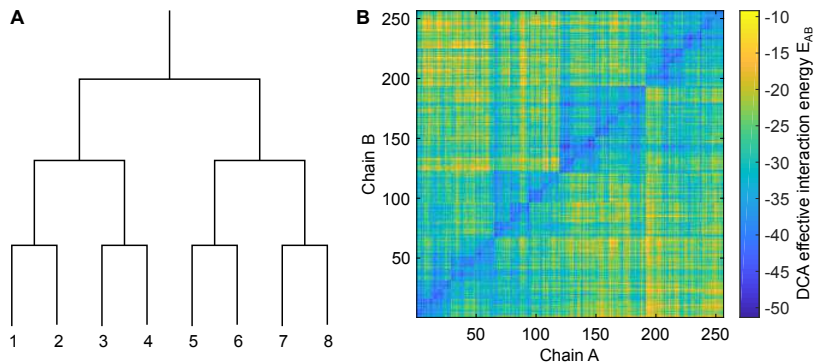}
	\vspace{0.2cm}
	\caption{{\bf Pairs of chains with common evolutionary history have small DCA effective interaction energies.} \textbf{A: } Chains are numbered according to the phylogenetic tree representing the branching process used for data generation (see Fig.~\ref{Fig1}). The same numbering is employed for chains A and for chains B that possess the same evolutionary history. \textbf{B: } Matrix of DCA effective interaction energies $E_{AB}$ (Eq.~\ref{energy}) for all pairs AB made from a chain A and a chain B of the testing set, numbered according to phylogeny as illustrated in panel A. Data was generated using a tree of 10 generations, with exactly 5 mutations per branch, out of 200 bits in each chain AB, thus yielding 1024 chains AB. Next, 75\% of them were randomly selected to form the training set employed to build the DCA model, while the remaining 25\% constitute the testing set.}
	\label{Fig2}
\end{figure}

Specifically, we generated data using a phylogenetic tree of 10 generations, with 5 mutations per branch, out of 100 bits in each complete chain AB, thus yielding 1024 chains AB. Given the relatively small number of mutations per branch, many of the resulting chains AB possess substantial similarities arising from their common evolutionary history. Specifically, Fig.~\ref{FigS2}A shows the histogram of Hamming distances between all B chains in the dataset, featuring a typical fraction 0.3 of sites with different states. The degree of similarity between two given chains arises from their relatedness along the phylogenetic tree used to generate the data. We ordered the chains employing this phylogenetic tree, so that sister chains are closest to one another etc. (see Fig.~\ref{Fig2}A).

Next, we randomly picked 75\% of the chains to form a substantial training set, and inferred a DCA model from this training set (see Methods). We employed the inferred couplings $e_{ij}(\sigma_i,\sigma_j)$ to compute effective interaction energies $E_{AB}$ (Eq.~\ref{energy}) between all chains A and all chains B of the remaining 25\% of the dataset, which constitutes our testing set. The effective interaction energies obtained are shown in Fig.~\ref{Fig2}B. Importantly, the diagonal of the matrix, corresponding to actual evolutionary partners, features small energies. Furthermore, a nested block structure is apparent in the matrix, reflecting the phylogenetic tree (recall that chains A and B are both ordered according to the tree as shown in Fig.~\ref{Fig2}A). Specifically, for 22\% of chains A in the testing set, the smallest DCA effective interaction energy $E_{AB}$ is obtained with their evolutionary partner (corresponding to the diagonal in Fig.~\ref{Fig2}B). In Fig.~\ref{FigS2}, we further demonstrate that those chains B that have smaller $E_{AB}$ with a chain A than its evolutionary partner B are very similar to that chain B and strongly related to it. Furthermore, if the dataset is divided into random species with 4 chains AB each (see Methods), the smallest $E_{AB}$ for a chain A within its species accurately identifies its evolutionary partner B for 93\% of chains A of the testing set. Hence, with a large training set, DCA is able to learn phylogenetic correlations, and to identify evolutionary partners. Recall that the usual goal of DCA is to infer couplings $e_{ij}(\sigma_i,\sigma_j)$  stemming from actual interactions, which do not exist in our synthetic data. 

\subsection*{Impact of key parameters and comparison to other methods}

\begin{figure}[h!]
	\centering
	\includegraphics[width=\textwidth]{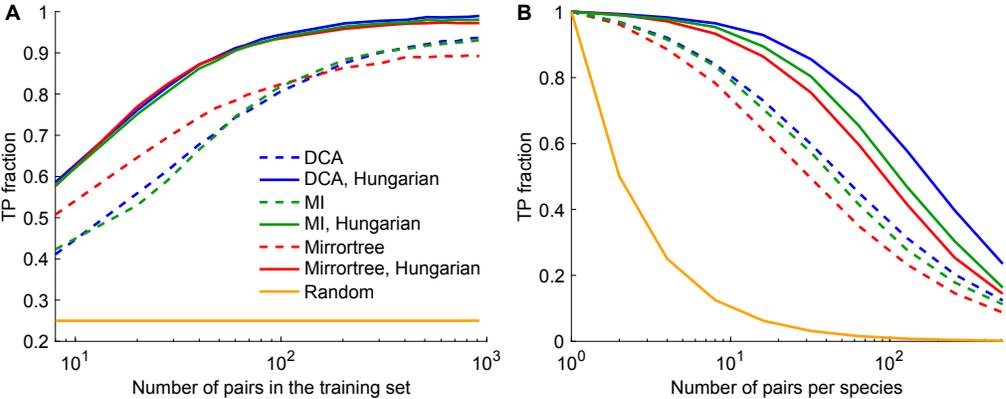}
	\vspace{0.2cm}
	\caption{{\bf Performance of pairing prediction versus training set size and number of pairs per species.} 
	 \textbf{A: } Fraction of pairs correctly identified (TP fraction) versus training set size, for DCA-, MI-, and Mirrortree-based methods. The three pairing scores corresponding to each of these three methods are employed in two ways: either within each species we find the chain B with optimal pairing score with each chain A (dashed lines), or within each species we employ the Hungarian matching algorithm to find the one-to-one pairing of chains A and B that optimizes the sum of the pairing scores (solid lines). Each species comprises 4 chains AB.  \textbf{B: } Fraction of pairs correctly identified (TP fraction) versus number of pairs per species, employing the same methods (and same colors) as in panel A, and a training set of 50\% of the total dataset. In both panels, data was generated using a tree of 10 generations, with exactly 5 mutations per branch, out of 200 bits in each chain AB, thus yielding 1024 chains AB. Species were built randomly, and some of them were chosen randomly to build the training set, the remaining ones making up the testing set. Yellow curves show the chance expectation, i.e. the average TP fraction obtained for random within-species pairings. Results are averaged over 100 replicates in panel A and 20 replicates in panel B, each corresponding to a different realization of the branching process used for data generation. The standard deviation of the TP fraction is 2-3\% for large training sets with 4 chains per species (panel A).}
	\label{Fig3}
\end{figure}

Let us investigate the robustness of the ability of DCA to identify pairs of evolutionary partners, and compare it to other methods. First, we ask how large a training set is necessary to learn the correlations arising from phylogeny. Fig.~\ref{Fig3}A shows that a sufficiently large training set is required for DCA to accurately identify evolutionary partners within each species, in line with previous results about DCA-based predictions of protein-protein interactions~\cite{Bitbol16,Gueudre16} and three-dimensional protein structures~\cite{Weigt09,Marks11,Morcos11} from real sequences. Furthermore, similar trends are observed both when employing the Mutual Information (MI) based score $S_{AB}$ (Eq.~\ref{SAB}), consistently with~\cite{BitbolPMI}, and when using the Mirrortree-inspired score $M_{AB}$ (Eq.~\ref{MAB}) that only relies on sequence similarity. All these methods predict pairings much better than the chance expectation (yellow) and reach very high fractions of true positives for training sets larger than $\sim$100 pairs AB (see Fig.~\ref{Fig3}A). Better performance is obtained when pairings are predicted using the Hungarian algorithm, which finds a global optimal one-to-one matching within each species (see Methods) than when simply picking for each chain A the optimal partner B within its species~\cite{BitbolPMI}. Fig.~\ref{Fig3}A further shows that with the first approach, the Mirrortree method performs better for small training sets, while DCA and MI outperform it for larger datasets. These differences become almost negligible when using the Hungarian algorithm.

Since the pairing task becomes harder when the number of pairs per species increases, we next studied how performance is affected by this important parameter. Fig.~\ref{Fig3}B, which employs a substantial training set, shows that the performance of all three pairing scores decays as species contain more pairs AB, as expected. However, this decay is far slower than for the chance expectation (yellow), which highlights the robustness of our methods. Here, DCA reaches the highest performance, followed by MI and then by Mirrortree, in line with the results obtained on Fig.~\ref{Fig3}A for large training sets. The good performance of the Mirrortree approach, which just relies on sequence similarities, arises from the fact that a possible pair AB that is very similar to correct pairs tends to be a correct pair too, as evidenced in Fig.~\ref{FigS3}. Indeed, pairs that are very similar to correct pairs tend to be their close ``relatives'' along the tree. Other variants based on sequence similarity can thus be constructed. In Fig.~\ref{FigS4}, we present two such variants: one employs as a pairing score the smallest Hamming distance from a possible pair AB of the testing set to its closest neighbor in the training set, and the second one is based on the notion of orthologous pairs. Both of them perform very well with large training sets, but are less robust than our other methods to decreasing training set size.

Because the ability of our methods to predict pairings relies on the shared evolutionary history of chains A and B, it is crucial to understand how the mutation rate affects performance. Let $\mu=n\mu_0$ denote the average total number of mutations between the ancestral complete chain AB and any complete chain AB at a leaf of the phylogenetic tree, where $n$ represents the number of generations and $\mu_0$ the average number of mutations per generation (see Fig.~\ref{Fig1}). If the maximum number of differences between two complete final chains AB, namely $2\mu$, becomes larger than the total length $2L$ of a complete chain AB, then correlations are lost between these two least-related chains AB. Thus, we expect the performance of our pairing prediction methods to decay for $\mu\gtrsim L$. This constitutes a lower bound of the actual number $\mu$ of mutations causing performance to substantially drop, because (i) we have considered the two least-related chains along the trees, and chains that diverged upon later duplication steps are more correlated, and (ii) since each mutation affects a random site, and each site can mutate several times, some sites may never mutate even when $\mu\gtrsim L$, and thus some correlations can survive in this regime even between the most distant sequences. Similarly, if $\mu_0\gtrsim L$, i.e. $\mu\gtrsim nL$, then even sister complete final chains AB  lose correlation, which gives an upper bound for the number of mutations causing performance to drop. 

\begin{figure}[h!]
	\centering
	\includegraphics[width=\textwidth]{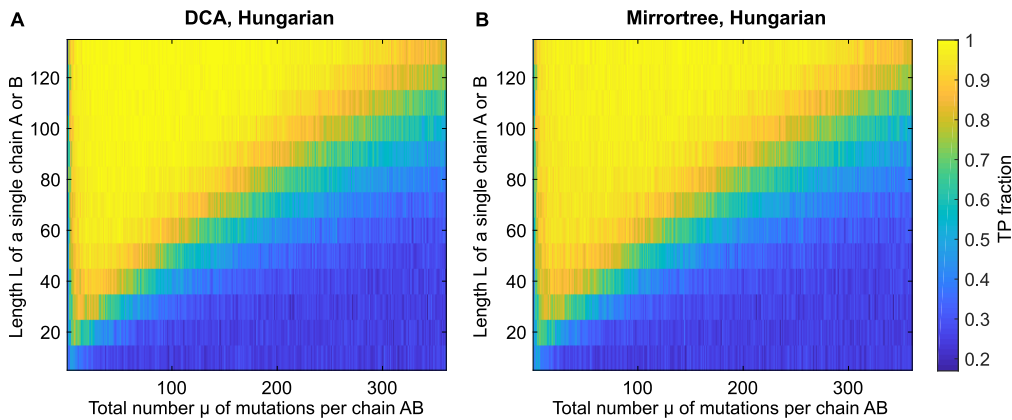}
	\vspace{0.2cm}
	\caption{{\bf Performance of DCA- and Mirrortree-based predictions for various data parameters.} The fraction of pairs AB correctly predicted (TP fraction) is shown versus the average total number of mutations $\mu$ per chain AB and the length $L$ of a single chain A or B for DCA (panel \textbf{A}) and Mirrortree (panel \textbf{B}). The Hungarian algorithm was employed to predict pairings. For each $\mu$ and $L$, data was generated using a tree of $n=10$ generations, thus yielding 1024 chains AB, and random species, each comprising 4 chains AB, were constructed. Half of the species were chosen to form a training set of 512 pairs, and predictions were made on the remaining species, which form the testing set. Here the chance expectation of TP fraction, obtained for random within-species pairings, is 0.25.}
	\label{Fig4}
\end{figure}

Fig.~\ref{Fig4} shows heatmaps of the performance of DCA- and Mirrortree-based pairing predictions versus the total number $\mu$ of mutations per chain AB and the single chain length $L$, with a substantial training set. For both methods, performance is very good and robust over a large range of values of $L$ and $\mu$. In addition, in both cases, a clear transition between good and poor performance is visible as $\mu$ is increased at each $L$. We observe that this transition occurs along a line, such that good performance is obtained for $\mu\lesssim 3.6\,L - 72$. This linear behavior and its slope are consistent with our predictions above. We also observe that performance drops if there are extremely few mutations, because chains are too conserved and remain almost all the same, and if chains are too short, because there is too much redundancy. Another important parameter is the total number $n$ of generations in the phylogenetic tree, which sets the total number of chains ($2^n$). We found that varying $n$ from 8 to 12 yielded no significant change the heatmaps of Fig.~\ref{Fig4}. Finally, the DCA- and Mirrortree-based methods perform very similarly over the whole range of parameters studied in Fig.~\ref{Fig4}: specifically, the mean difference of the TP fractions obtained using the two methods is $4\times 10^{-3}$. Despite their conceptual differences, these methods rely on learning phylogenetic correlations from the training set, and thus have similar dependences on evolutionary parameters such as the mutation rate. 

\subsection*{Evolutionary partners can be predicted without a training set}

In Refs.~\cite{Bitbol16,BitbolPMI}, it was shown that DCA- and MI-based approaches allow to predict interacting partners among the paralogs of actual interacting proteins from their sequences without any training set, i.e. without any prior knowledge of interacting pairs, thanks to an Iterative Pairing Algorithm (IPA). In this approach, at the first iteration, (usually poor) predictions are made employing pairing scores learned on random within-species pairings. At each subsequent iteration $n > 1$, the predictions from the previous iteration that are deemed most reliable~\cite{Bitbol16,BitbolPMI} are progressively incorporated. Specifically, pairing scores are re-learned on the $(n-1)N_\mathrm{increment}$ top-ranked predicted pairs from the previous iteration, where $N_\mathrm{increment}$ represents the increment step: hence, the number of predicted pairs that are employed to make the next predictions increases by $N_\mathrm{increment}$ at each iteration (see Refs.~\cite{Bitbol16,BitbolPMI} for details, and Ref.~\cite{Gueudre16} for alternative iterative approaches). This iterative strategy gradually improves predictive power and has yielded accurate predictions of interacting partners in ubiquitous prokaryotic protein families~\cite{Bitbol16,BitbolPMI}. Here, we employed the IPA on our synthetic data where correlations arise only from shared evolutionary history. For comparison, we also developed and studied a variant of the IPA based on the Mirrortree approach, which employs the pairing score $M_{AB}$ (Eq.~\ref{MAB}), instead of the DCA-based effective interaction energy $E_{AB}$ (Eq.~\ref{energy})~\cite{Bitbol16} or MI score $S_{AB}$ (Eq.~\ref{SAB})~\cite{BitbolPMI}.

\begin{figure}[h!]
	\centering
	\includegraphics[width=\textwidth]{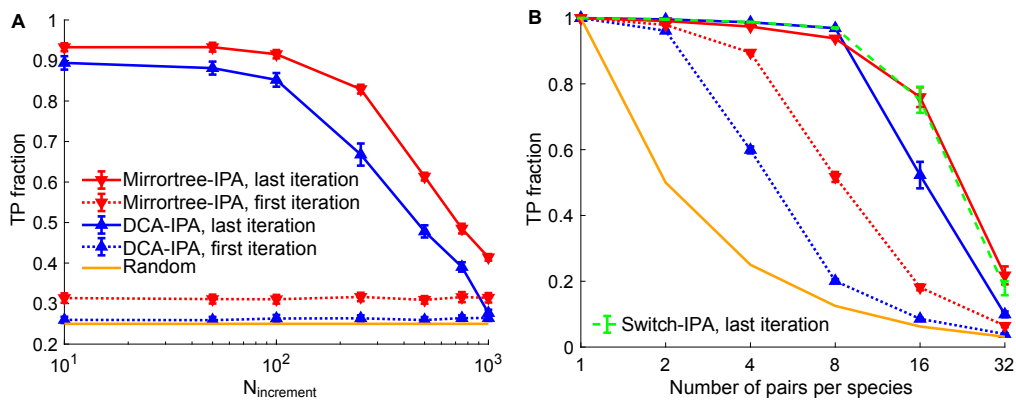}
	\vspace{0.2cm}
	\caption{{\bf Pairing prediction without any training set.} \textbf{A:} The fraction of pairs AB correctly predicted (TP fraction) is shown versus the increment step $N_\textrm{increment}$ of the iterative process, for the DCA-IPA and the Mirrortree-IPA, at the first and last iterations. Data was generated using a tree of 10 generations, with 20 mutations per branch on average, out of 200 bits in each chain AB, thus yielding 1024 chains AB, and random species with 4 pairs AB each were constructed. \textbf{B:} TP fraction versus number of pairs per species, for the DCA-IPA and the Mirrortree-IPA, at the first and last iterations, as well as for the Switch-IPA, which uses the Mirrortree pairing score for the first half of iterations and then switches to the DCA pairing score (the first iteration is thus the same for the Switch-IPA as for the Mirrortree-IPA). An increment step $N_\textrm{increment}=100$ was used. Data was generated using a tree of 10 generations, with 5 mutations per branch on average, out of 200 bits in each chain AB, thus yielding 1024 chains AB, and random species with the same number of chains AB each were constructed. Colors are the same as in panel A, with the addition of the Switch-IPA. In both panels, predictions were made without any training set, and the first iteration employed random within-species pairings to compute the initial pairing scores. The Hungarian algorithm was employed to predict pairings. Results are averaged over 20 replicates in panel A and 100 in panel B, each corresponding to a different realization of the branching process used for data generation. Error bars represent 95\% confidence intervals (in many cases, markers are larger than error bars). }
	\label{Fig5}
\end{figure}

Fig~\ref{Fig5}A shows the TP fraction obtained for different values of the increment step $N_\mathrm{increment}$, both for the DCA-IPA and for the Mirrortree-IPA, at the first and last iterations. Overall, it shows that the iterative approach allows to make very accurate pairing predictions in the absence of a training set. With both algorithms, a strong improvement of predictive power is observed at the last iteration, compared to the first iteration and to the random expectation. Furthermore, the iterative method performs best for small increment steps $N_\mathrm{increment}$, which highlights the interest of the iterative approach. We emphasize that the high final TP fractions are attained without any prior knowledge of pairings. As discussed in Refs.~\cite{Bitbol16, BitbolPMI}, an important ingredient for the IPA to bootstrap its way toward high predictive power is that among pairs AB comprising a chain A and a chain B from the same species, correct pairs of partners possess more neighbors in terms of sequence similarity, quantified by the Hamming distance, than incorrect pairs. Ref.~\cite{Bitbol16} called this the \textit{Anna Karenina effect}, referring to the first sentence of Tolstoy's novel. We studied the Anna Karenina effect in our synthetic dataset. Fig.~\ref{FigS3} shows that correct pairs have closer correct neighbors than incorrect pairs: for instance, employing a threshold Hamming distance of 0.15 to define neighbors (see Fig.~\ref{FigS3}A),  correct pairs have 6.2 neighbors on average, of which 90\% are correct pairs, while incorrect pairs have 0.63 neighbors on average, of which 33\% are correct pairs (see Fig.~\ref{FigS3}B). In this case, correct pairs thus have almost 10 times more neighbors than incorrect ones, demonstrating a strong Anna Karenina effect. This favors correct pairs in the IPA, especially at early iterations~\cite{Bitbol16}.

For the parameters used in Fig.~\ref{Fig5}A, the Mirrortree-IPA performs slightly better than the DCA-IPA, and this difference exists right from the first iteration. How does the performance of these two methods depend on parameters characterizing the dataset? Fig.~\ref{FigS5} shows heatmaps of the performance of the DCA-IPA and of the Mirrortree-IPA as a function of the total number $\mu$ of mutations per chain AB and of the single chain length $L$, without any training set, at the first and last iterations. It shows that the Mirrortree-IPA performs better than the DCA-IPA at the first iteration, especially as mutation rates become larger (Fig.~\ref{FigS5}A and C). Recall that at the first iteration, pairing scores are calculated on random within-species pairings, where most pairs (75\% on average for species with 4 pairs each) are incorrect. Taken together with our earlier observation that Mirrortree outperforms DCA for small training sets (see Fig.~\ref{Fig3}A), it means that DCA requires a substantial and accurate training set to properly learn correlations and reach good performance, as is already known in the case of real protein sequences~\cite{Weigt09, Marks11,Morcos11}. Nevertheless, the IPA allows DCA to robustly reach high predictive power at the last iteration over a broad range of values of $\mu$ and $L$ (Fig.~\ref{FigS5}B). This range is almost as large as for the Mirrortree-IPA (Fig.~\ref{FigS5}D), despite the initial lower performance of DCA. At the last iteration, there is only a rather narrow band, close to the transition line from large to small TP fractions, where the DCA-IPA is outperformed by the Mirrortree-IPA. Note that the parameters employed in Fig.~\ref{Fig5}A are in this region. Importantly, the parameter range where the DCA-IPA and the Mirrortree-IPA perform well without any training set is very similar to that obtained in Fig.~\ref{Fig4} for DCA and Mirrortree predictions from a large training set. This result illustrates the power of the iterative approach, which truly allows to bypass the need for a training set.

In Fig.~\ref{Fig5}B, we further investigate the impact of the number of pairs AB per species on performance of the DCA-IPA and of the Mirrortree-IPA without any training set. Very good performance is obtained for species comprising up to 8 pairs AB, and then we observe a decay, which is steeper than with a large training set (Fig.~\ref{Fig4}). We further observe that the DCA-IPA reaches higher performance than the Mirrortree-IPA for small numbers of pairs per species, while the opposite is true for larger numbers of pairs per species. Again, the Mirrortree-IPA performs better than the DCA-IPA at the first iteration, and this might explain the decreased final performance of the DCA-IPA for large numbers of pairs per species: confronted with many pairing possibilities, the DCA-IPA may not be able to recover from the large amount of noise in the initial matches.

An interesting question is whether one can improve predictive power by combining DCA and Mirrotree approaches. It is all the more attractive that the predictions made using the two scores are almost independent (Fig.~\ref{FigS6}A), in contrast to those made using DCA and MI~\cite{BitbolPMI} (Fig.~\ref{FigS6}B). Because Mirrortree performs better than DCA at the first iteration while DCA becomes better for larger and more accurate training sets, we devised a version of the IPA that uses the Mirrortree pairing score for the first half of iterations and then switches to the DCA pairing score. The final TP fractions obtained with this Switch-IPA are shown in Fig.~\ref{Fig5}B. We find that it performs as well as the best among the Mirrortree-IPA and the DCA-IPA, which should make it more broadly applicable. However, it does not yield further improvements.

\subsection*{Extension to other phylogenies and application to real proteins}

So far, we have considered a minimal model for species, where chains are randomly grouped into sets of equal size $m$, each representing a species. Such a model would be realistic in the case where exchange between species (i.e. horizontal gene transfer) is very frequent. But in other evolutionary regimes, different correlations are expected between the chains present in a given species (paralogs) and between the most closely related chains across species (orthologs), due to the fact that species are evolutionary units. In practice, paralogous and orthologous pairs respectively arise from duplication and speciation events. In a duplication event, a chain from a given species gives rise to 2 paralogous chains within this species. Note that loss events can also occur, thus decreasing the number of chains within a species. In a speciation event, all chains are duplicated to give rise to 2 distinct species. In order to assess the robustness of our results to these effects, we now consider a phylogeny model that explicitly accounts for duplication-loss and speciation events, without any exchange within species (see Fig.~\ref{FigS7}A). For simplicity, we assume that duplication and loss always happen together, so that the number of pairs per species remains constant. 

Fig.~\ref{FigS7}B shows the performance of DCA and Mirrortree scores at predicting pairs of evolutionary partners from a substantial training set, versus the fraction of species that undergo a duplication-loss event upon speciation. Overall, performance is good, but it decreases when the frequency of duplication-loss events increases. When there are no duplication-losses, this model features $m$ distinct phylogenies ($m=4$ in Fig.~\ref{FigS7}), one for each ancestral chain, and similarities between the chains of the testing set and of the training set that belong to the same phylogeny allow to predict evolutionary partners. Conversely, when there are many duplication-losses, chains from one single phylogeny will end up fixing in each species, analogously to asexual birth-death population genetics models at fixed population size where all individuals are descended from a single ancestor after a sufficient time~\cite{Ewens79}. Moreover, chains resulting from recent duplication events will be very hard to distinguish, resulting in pairing ambiguities. Fig.~\ref{FigS7}B also shows that DCA outperforms Mirrortree, consistently with other cases with a substantial training set. Finally, in Fig.~\ref{FigS7}C, we consider the case without a training set. We find that the first iteration of the Mirrortree-IPA performs much better than that of the DCA-IPA, while things are more even after the iterative process, consistently with our previous results. Furthermore, Fig.~\ref{FigS7}C shows that in the absence of duplication-loss, final performance is close to the random expectation of 0.25, which stands in contrast with the case with a training set. This is because in the absence of a training set, incorrect pairs comprising A and B chains from different phylogenies cannot be distinguished from those coming from the same phylogeny. Duplication-loss events allow to break this symmetry thanks to the fixation process whereby possible cross-phylogeny pairs become rare within each species. However, when duplication-loss events are too frequent, performance decays again, for the same reason as with a training set: pairs resulting from recent duplications are hard to distinguish. This tradeoff yields an optimal performance for an intermediate frequency of duplication-loss events. Overall, our DCA and Mirrortree-based pairing predictions are robust to modifications of the phylogeny used to generate the data, as long as evolutionary correlations exist.

\begin{figure}[h!]
	\centering
	\includegraphics[width=\textwidth]{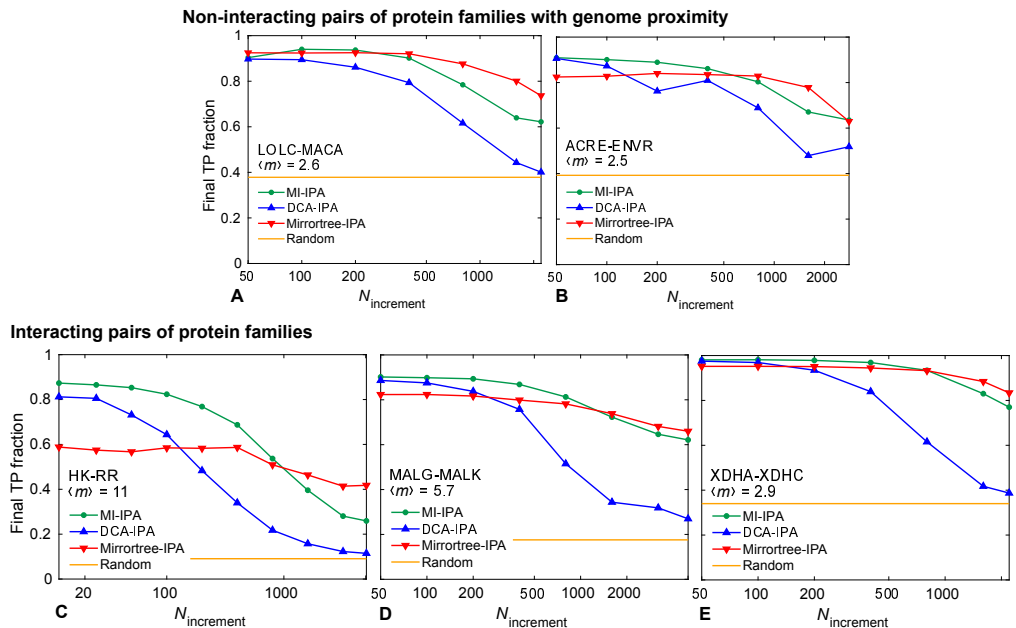}
	\vspace{0.2cm}
	\caption{{\bf Pairing predictions for real pairs of protein families.} The final fraction of protein pairs correctly predicted (TP fraction) obtained without a training set by the MI-IPA, the DCA-IPA and the Mirrortree-IPA is shown versus the increment step $N_{\mathrm{increment}}$ of the iterative process. \textbf{A, B:} Pairs of protein families with no known direct physical interaction but that are encoded closely on the genome; datasets include $\sim 2000$ homologous pairs. \textbf{C, D, E:} Pairs of protein families with known direct physical interactions; for large families (C and D), datasets of $\sim 5000$ pairs comprising only full species were extracted from the larger complete datasets, while in E, the full dataset of $\sim 2000$ pairs was used. In all panels, the mean number $\langle m\rangle$ of pairs per species is indicated, and yellow lines represent the average TP fraction obtained for random within-species pairings. All results are averaged over 50 replicates that differ in their initial random within-species pairings. Datasets were constructed as described in~\cite{Bitbol16,BitbolPMI}, starting from the P2CS database~\cite{Barakat09,Ortet15} for histidine kinase-response regulator (HK-RR) dataset (C) and using a method adapted from~\cite{Ovchinnikov14} that relies on finding homologs of \textit{Escherichia coli} protein pairs in all other cases.}
	\label{Fig6}
\end{figure}

Since the various methods presented here allow to reliably pair synthetic chains on the basis of shared evolutionary history only, it should also be the case for real proteins. While it is difficult to be certain that two real protein families share evolutionary history but do not bear common functional evolutionary constraints, we chose two pairs of protein families that are generally encoded in close proximity on prokaryotic genomes but that do not have known direct physical interactions~\cite{Szklarczyk19}, namely the \textit{Escherichia coli} protein pairs LOLC-MACA and ACRE-ENVR and their homologs. (Note that ENVR has regulatory roles on ACRE expression~\cite{Hirakawa08}.) Indeed, we expect proteins encoded close to one another to share common evolutionary history, because they tend to be horizontally transferred together, and to have similar levels of expression and evolutionary rates~\cite{Hakes07,Swapna12}. Fig.~\ref{Fig6}A and B shows that the DCA-, MI- and Mirrortree-IPA are all able to reliably pair LOLC-MACA and ACRE-ENVR homologs that are encoded close to one another on genomes, despite the absence of known direct physical interactions between these protein families. We further compared these results to three pairs of protein families with known direct physical interactions~\cite{BitbolPMI}. Fig.~\ref{Fig6}C, D and E shows that the Mirrortree-IPA performs less well than the DCA and MI-based versions in these cases, especially for the dataset of histidine kinases and response regulators (Fig.~\ref{Fig6}C). These proteins feature a strongly coevolving interaction interface~\cite{Laub07}, diversified across many paralogs per species (average number of pairs per species in our dataset: $\left<m\right>=11$) to avoid unwanted crosstalk. This argues in the favor of the DCA-IPA and the MI-IPA rather than Mirrortree-based methods in order to predict partnership among physically interacting partners. Interestingly, the MI-IPA slightly outperforms the DCA-IPA for all the real datasets in Fig.~\ref{Fig6}, while this is not the case for synthetic data comprising only phylogenetic correlations (see Fig.~\ref{Fig3}). 

\newpage

\subsection*{Distinguishing physically interacting proteins from proteins only sharing evolutionary history}

We have shown that the IPA, which aims to find partners among the paralogs of two protein families, accurately identifies pairs of proteins even if they only share a common evolutionary history, in the absence of direct interactions. In order to computationally predict protein-protein interactions from sequence data, it is in addition necessary to distinguish the pairs of protein families that physically interact from those that only share evolutionary history.

First, interacting protein families often share more evolutionary history than non-interacting ones. This idea is at the heart of the Mirrortree approach~\cite{Pazos01,Ochoa10,Ochoa15}, which is usually implemented on alignments of orthologs, thus allowing to predict which protein families interact, but not addressing the paralog pairing problem. We have applied the original Mirrortree~\cite{Pazos01,Ochoa10} and pMirrortree~\cite{Ochoa15} methods to the pairs of protein families studied in Fig.~\ref{Fig6}, focusing on the orthologs of the reference \textit{E. coli} protein pairs (see Table~\ref{TableS1}). The two pairs without known physical interactions (LOLC-MACA and ACRE-ENVR) feature the smallest Mirrortree scores among the five pairs considered, and they also have non-significant pMirrortree scores, while two out of the three pairs with known direct physical interactions (HK-RR and XDHA-XDHC) possess significant pMirrortree scores. These rather encouraging results confirm the power of the Mirrortree method, and support the hypothesis that our two pairs without known physical interactions are really non-interacting.

\begin{figure}[h!]
	\centering
	\includegraphics[width=0.6\textwidth]{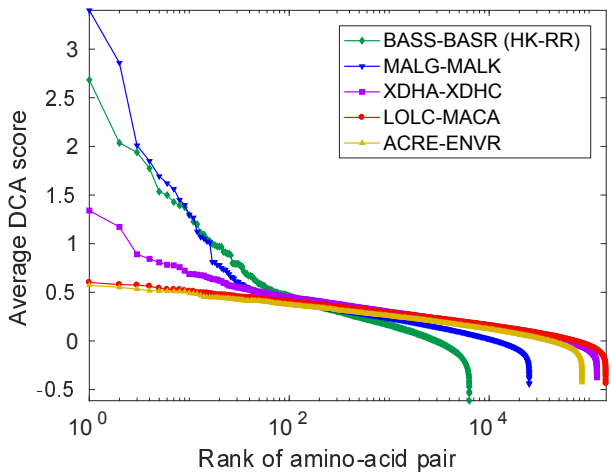}
	\vspace{0.2cm}
	\caption{{\bf Outliers in DCA score provide evidence of protein-protein interactions.} DCA scores (APC-corrected Frobenius norms~\cite{Ekeberg13,Ovchinnikov14}) were evaluated for each pair of amino-acid sites at the final IPA iteration. Their averages over 500 IPA replicates that differ in their initial random pairings of sequences are shown ranked by decreasing value. Datasets correspond to the same protein family pairs as in Fig.~\ref{Fig6}, and the IPA was run with $N_\mathrm{increment} = 50$. Outliers in DCA score appear for all three pairs of protein families with known direct physical interactions (BASS-BASR, MALG-MALK, XDHA-XDHC), but not for the other ones (LOLC-MACA, ACRE-ENVR).}
	\label{Fig7}
\end{figure}

Another way to distinguish physically interacting pairs of protein families from non-interacting ones is to leverage the DCA scores of amino-acid pairs. Indeed, strong DCA scores tend to correspond to contacting amino-acid pairs~\cite{Weigt09,Morcos11,Marks11}, and thus, their presence can reveal actual interacting partners~\cite{Ovchinnikov14,Feinauer16,Bitbol16}. Fig.~\ref{Fig7} demonstrates that outliers in DCA scores~\cite{Bitbol16} exist for the three pairs with known direct physical interactions studied in Fig.~\ref{Fig6} (HK-RR, MALG-MALK and XDHA-XDHC), while they are absent for the two pairs without known physical interactions (LOLC-MACA and ACRE-ENVR). Therefore, in these examples, outliers in DCA scores successfully allow to distinguish pairs of protein families that physically interact from those that only share a similar evolutionary history. Importantly, the results on Fig.~\ref{Fig7} were obtained on the pairs of protein sequences predicted by the IPA, which means that this method can be directly combined with the IPA. Note that the absence of outliers in DCA scores in the pairs without known physical interactions hints that phylogenetic correlations result in multiple small DCA couplings, while direct physical interactions yield few large DCA couplings. We also checked that our phylogeny-only synthetic data (with the same parameters as in Fig.~\ref{Fig2}) does not yield outliers in DCA scores.

Therefore, combining either traditional Mirrortree approaches or the study of outliers in DCA scores with the IPA can allow both to predict interacting pairs of protein families and to solve the paralog pairing problem, starting from two independent multiple sequence alignments of the protein families.

\newpage

\section*{Discussion}

Recently, methods relying on pairwise maximum entropy DCA models, originally employed to identify amino acids in contact in the three-dimensional structure of proteins and multi-protein complexes, have allowed to reliably predict interacting partners among the paralogs of several ubiquitous prokaryotic protein families, starting from sequences only~\cite{Bitbol16, Gueudre16}. An important motivation for these methods is that the need to maintain the physico-chemical complementarity of contacting amino acids induces correlations in amino-acid usage between the sequences of interacting partners~\cite{Laub07,Weigt09}. However, correlations between the sequences of interacting partners can also arise from their shared evolutionary history~\cite{Lovell10}. 

In the present work, employing controlled synthetic data, we demonstrated that DCA is able to accurately identify partners that only share evolutionary history, in the absence of functional and structural constraints. This result holds even in the absence of a training set, thanks to the Iterative Pairing Algorithm (IPA)~\cite{Bitbol16}. Because our controlled synthetic data only comprises signal from phylogeny, we compared our DCA-based approach to methods that explicitly rely on phylogeny, through sequence similarity. Specifically, we proposed a method based on the Mirrortree approach~\cite{Pazos01,Ochoa10,Ochoa15} to predict pairs among the paralogs of two protein families. We obtained similar performances, with DCA slightly outperforming Mirrortree when a substantial correct training set is available. We also considered a method based on Mutual Information (MI)~\cite{BitbolPMI}, yielding similar performances for our synthetic data, with DCA often slightly outperforming MI, while MI tends to outperform DCA for natural sequence data~\cite{BitbolPMI}. The robustness of MI to finite dataset size effects~\cite{BitbolPMI} and its ability to quantify statistical dependence whatever its origin might make it most appropriate for complex natural data. 

Finally, we applied the DCA-IPA, the MI-IPA and the Mirrortree-IPA to natural sequence data from several pairs of prokaryotic protein families, with or without known direct physical interactions, but always with genome proximity and thus significant shared evolutionary history. We obtained accurate predictions for all these datasets. Therefore, correlations from evolutionary history can play an important part in the performance of these algorithms in the case of natural sequence data. Interestingly, we found that the Mirrortree-IPA performs significantly less well than the MI-IPA and DCA-IPA on the histidine kinase-response regulator dataset, which features large numbers of paralogs per species and is known to possess strongly coevolving contacts~\cite{Laub07,Weigt09}, and to which DCA is thus particularly well-suited. This points to the complementarity of these methods. In addition, we showed that pairs of protein families with known direct physical interactions can be distinguished from those without known direct physical interactions, either by employing the original Mirrortree approach involving orthologs only~\cite{Pazos01,Ochoa10,Ochoa15}, or by studying the amino-acid pairs that are outliers in DCA score~\cite{Bitbol16}. Hence, combining these methods with the IPA allows to both predict interacting protein families and to find interacting partners among paralogs.

The ability of DCA to identify evolutionary partners in the absence of functional and structural constraints can be viewed as surprising. Indeed, DCA models are mainly known for their ability to identify structural contacts, and they emphasize small-eigenvalue modes of the covariance matrix of sequences~\cite{Rivoire13,Cocco13}, as illustrated by the inversion of the covariance matrix involved in the mean-field approximation (see Methods), while important signal from phylogeny lies in the large-eigenvalue modes of the covariance matrix~\cite{Casari95,Halabi09,Qin18}. In addition, phylogenetic correlations are often considered deleterious to structure prediction~\cite{Lapedes99,Weigt09,Marks11,Qin18}, which is one of the major applications of DCA. Nevertheless, a DCA model is fundamentally a global statistical model that aims to faithfully reproduce the empirical pairwise correlations observed in the data~\cite{Cocco18,Nguyen17}. As such, it should also encode phylogenetic correlations, thus rationalizing our result. Our findings demonstrate that DCA models capture coevolution in a broad sense, not limited to contacting pairs of residues, but also including shared evolutionary history unrelated to functional constraints. Importantly, while phylogenetic correlations are often viewed as a confounding factor for structure prediction~\cite{Lapedes99,Weigt09,Marks11,Qin18}, our work shows that they are actually useful in order to address the paralog pairing problem by DCA. This is consistent with previous work directly exploiting phylogenetic correlations to predict protein-protein interactions~\cite{Fryxell96,Goh00,Jothi05,Ochoa10,Ochoa15}. Interesting further directions include analyzing the contributions of phylogeny topology and evolutionary rates~\cite{Hakes07,Juan08,Kann09,Lovell10,Swapna12} in the phylogenetic signal that is useful for DCA-based pairing predictions, and addressing how signals from phylogeny and from amino-acid contacts combine together.

Here, our natural data examples were pairs of proteins colocalized on genomes, allowing us to easily evaluate performance via the fraction of correctly predicted protein pairs. However, computational interaction prediction and paralog pairing methods are particularly important in order to discover interaction partners that are not encoded in close genomic locations. This corresponds to the general case in eukaryotes, which makes this problem extremely relevant. In prokaryotes too, important interactions between very different cellular processes exist across operons~\cite{Peters16}. It will be of great interest to apply the IPA to such cases. One restrictive assumption we have made so far is that interactions are one-to-one, which is appropriate for strongly specific protein-protein interactions, but not for cases involving promiscuity, crosstalk, or non-interacting orphan proteins. Hence, an important direction for future work is to generalize the IPA to allow for multiple partners.

\section*{Acknowledgments}
AFB thanks Ned S. Wingreen and Yaakov Kleeorin for inspiring discussions. The authors thank Institut de Biologie Paris-Seine (IBPS) at Sorbonne Université for funding via a Collaborative Grant (Action Incitative) to AFB and MW. MW also acknowledges funding by the EU H2020 research and innovation program MSCA-RISE-2016 under grant agreement No. 734439 InferNet.

\newpage

\section*{Supporting information}
\def\theequation{S\arabic{equation}}
\setcounter{equation}{0}
\def\thefigure{S\arabic{figure}}
\def\thetable{S\arabic{table}}
\setcounter{figure}{0}


\begin{figure}[h!]
	\centering
	\includegraphics[width=\textwidth]{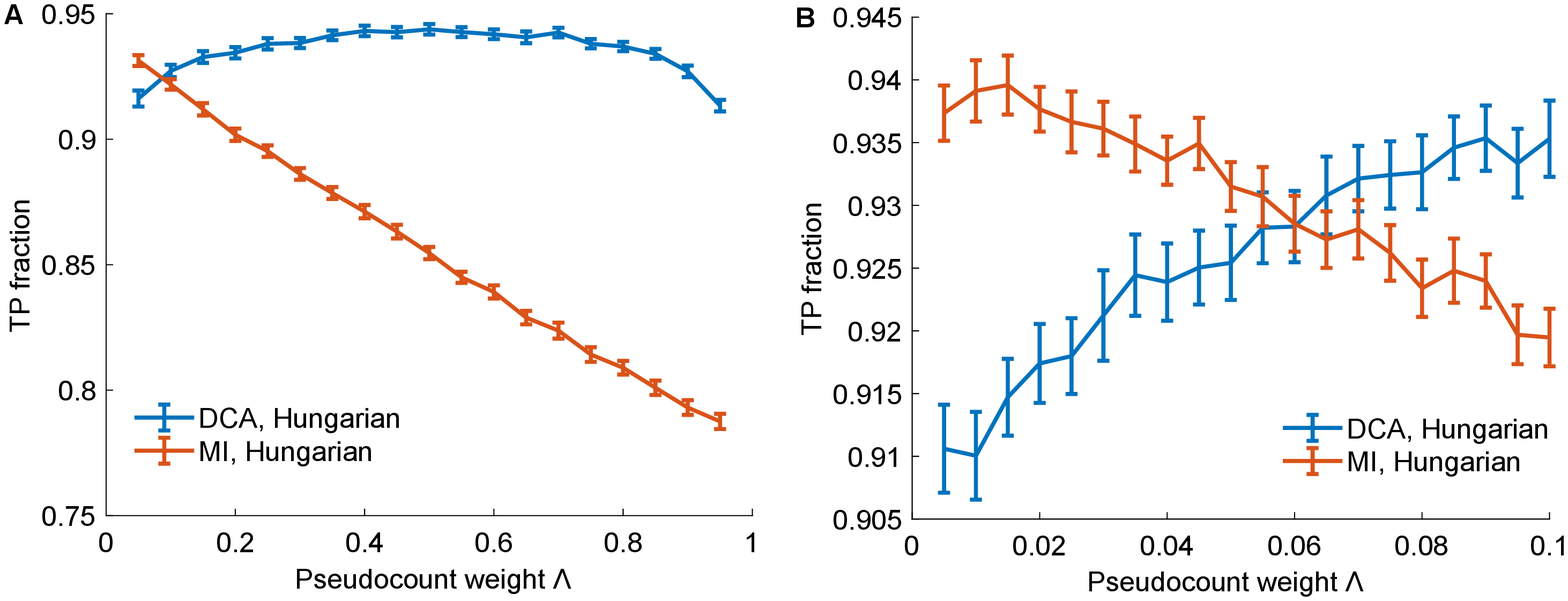}
	\vspace{0.2cm}
	\caption{{\bf Impact of the pseudocount weight on performance.} The fraction of pairs AB correctly predicted (TP fraction) is shown versus the pseudocount weight $\Lambda$ both for DCA- and for MI-based pairing methods. The Hungarian algorithm was employed to predict pairings. \textbf{A:} Full range of variation of $\Lambda$ (excluding $\Lambda=1$, where no more signal comes from the actual dataset, and accordingly, predictions follow the chance expectation of 0.25 TP fraction). \textbf{B:} Zoom over small values of $\Lambda$. In both panels, data was generated using a tree of 10 generations, with exactly 5 mutations per branch, out of 200 bits in each chain AB, thus yielding a total of 1024 chains AB. Random species with 4 chains AB each were constructed; 25 of them were randomly selected to form a training set of 100 pairs AB employed to build the DCA model, and the rest constitutes the testing set, for which we calculated DCA pairing scores (Eq.~\ref{energy}). Results are averaged over 100 realizations corresponding to different random partitions of the dataset into training and testing sets. Error bars represent 95\% confidence intervals.
}
	\label{FigS1}
\end{figure}

\begin{figure}[h!]
	\centering
	\includegraphics[width=\textwidth]{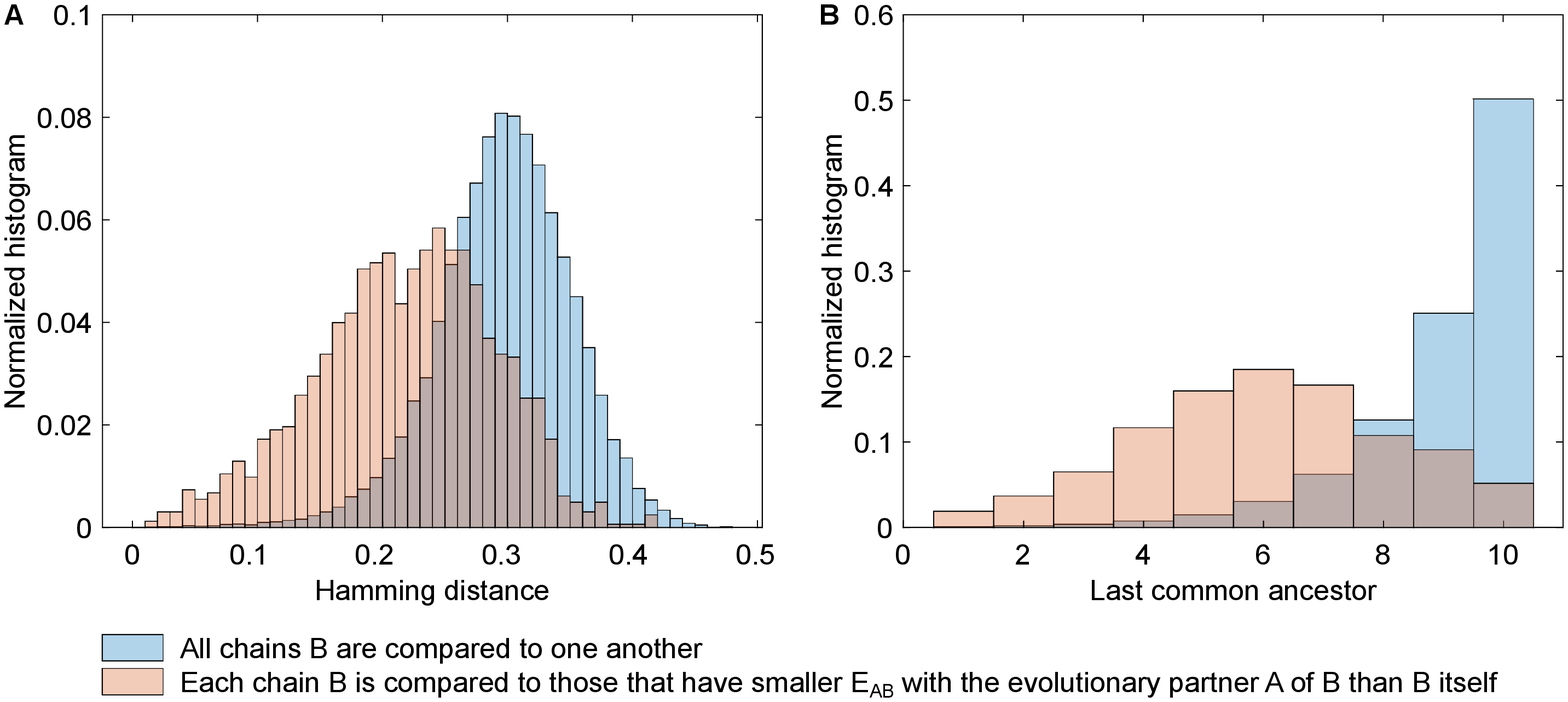}
	\vspace{0.2cm}
	\caption{{\bf Pairs with smaller DCA effective interaction energy than evolutionary partners are similar and closely related to them.} \textbf{A: } Blue: Histogram of Hamming distances between all chains B from the testing set. Red: Histogram of Hamming distances between each chain B from the testing set and those with smaller DCA effective interaction energy $E_{AB}$ (Eq.~\ref{energy}) than B itself with the evolutionary partner A of B. \textbf{B: } Blue: Histogram of the last common ancestor index between all chains B from the testing set. Red: Histogram of the last common ancestor index between each chain B from the testing set and those with smaller DCA effective interaction energy $E_{AB}$ (Eq.~\ref{energy}) than B itself with the evolutionary partner A of B. The last common ancestor index is defined according to the phylogenetic tree representing the branching process used for data generation (see Fig.~\ref{Fig1}): it is 1 for chains that are evolutionarily closest (``sisters''), 2 for the next level (``cousins''), etc. In both panels, data was generated using a tree of 10 generations, with exactly 5 mutations per branch, out of 200 bits in each chain AB, thus yielding 1024 chains AB. Next, 75\% of them were randomly selected to form the training set employed to build the DCA model and compute $E_{AB}$ values, while the remaining 25\% constitute the testing set. }
	\label{FigS2}
\end{figure}

\begin{figure}[h!]
	\centering
	\includegraphics[width=\textwidth]{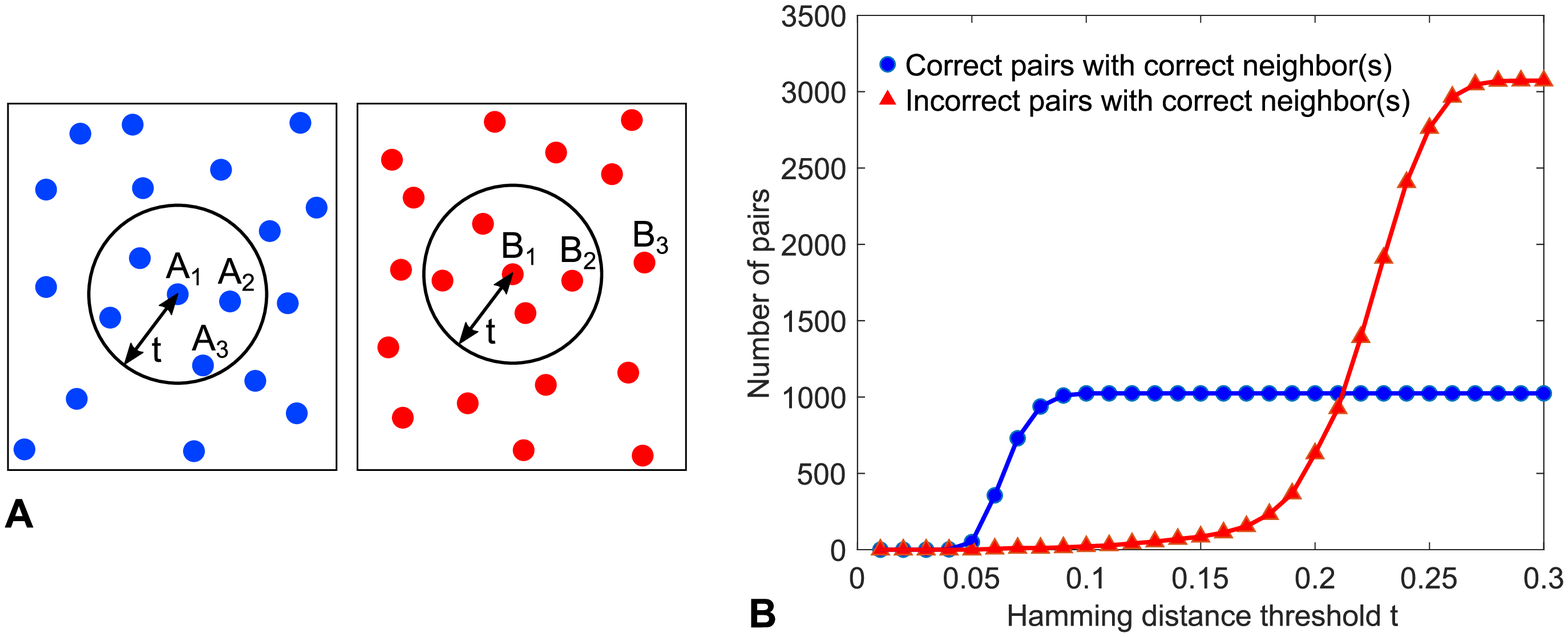}
	\vspace{0.2cm}
	\caption{{\bf Neighbors of correct pairs tend to be correct pairs.} \textbf{A: }Schematic illustrating the notion of neighbor pairs. On the left, each chain A is represented as a blue dot, and distances between dots correspond to Hamming distances, i.e. fractions of sites with different states. A threshold Hamming distance $t$ is chosen to define the notion of neighbors. The dots inside the circle with center $A_1$ and radius $t$ represent the neighbors of $A_1$. A similar schematic is shown on the right for the B chains. Two pairs $A_1B_1$ and $A_2B_2$ are considered neighbors if $A_1$ and $A_2$ are neighbors, and $B_1$ and $B_2$ are neighbors too. Thus, in our schematic, $A_2B_2$ is a neighbor of $A_1B_1$, but $A_3B_3$ is not. In practice, only pairs where A and B belong to the same species are considered, because they are the only possible correct pairs of evolutionary partners. \textbf{B: }The number of within-species pairs AB that possess neighbors that are correct pairs of evolutionary partners is shown versus the Hamming distance threshold $t$ defining the notion of neighbor (see panel \textbf{A}). The correct pairs, i.e. those are evolutionary partners, and the incorrect within-species pairs, made of a chain A and a chain B that are not evolutionary partners, are distinguished. For intermediate distance thresholds, there are far more correct pairs than incorrect pairs that possess correct neighbors. For instance, for a Hamming distance threshold of 0.1, 22 incorrect pairs have correct neighbors, while all 1024 correct pairs have correct neighbors. Data was generated using a tree of 10 generations, with exactly 5 mutations per branch, out of 200 bits in each chain AB, thus yielding 1024 chains AB, and random species with 4 chains AB each were constructed. }
	\label{FigS3}
\end{figure}

\begin{figure}[h!]
	\centering
	\includegraphics[width=\textwidth]{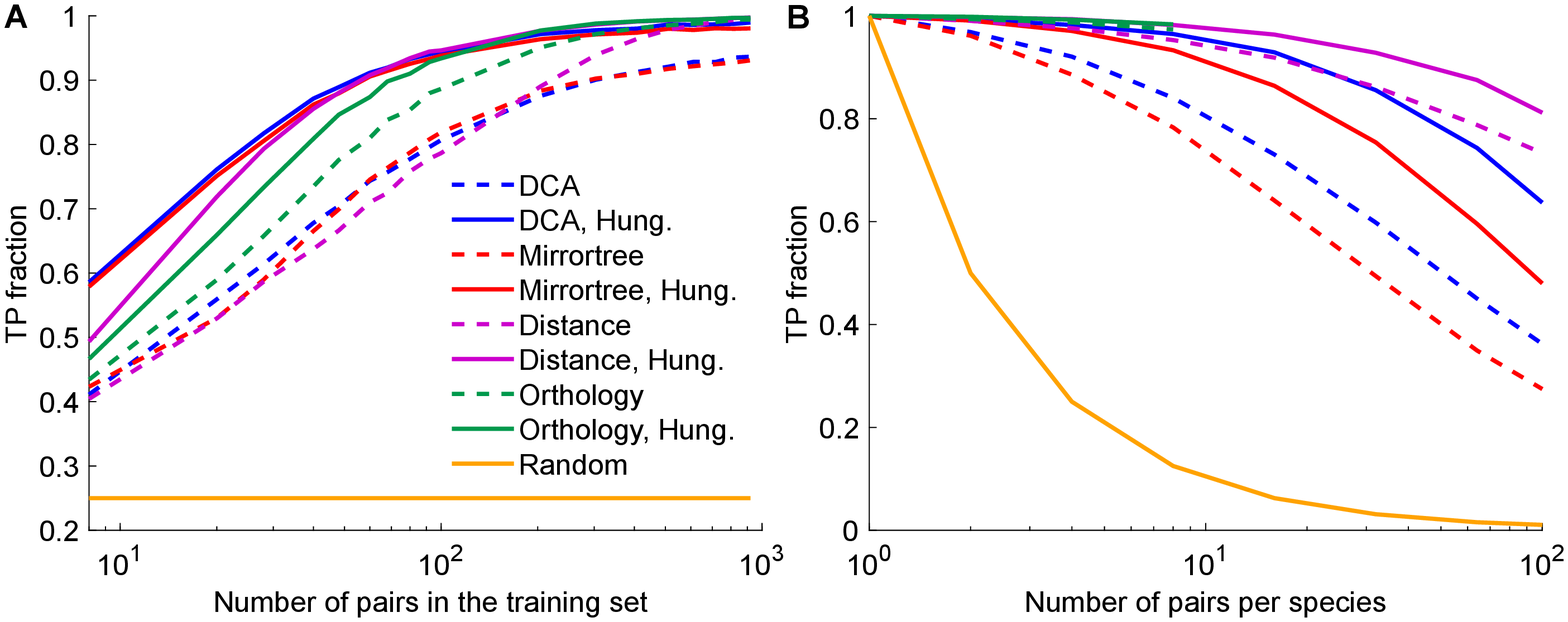}
	\vspace{0.2cm}
	\caption{{\bf Performance of pairing prediction versus training set size and number of pairs per species for various methods.} 
		\textbf{A: } Fraction of pairs correctly identified (TP fraction) versus training set size, for DCA- and MI-based methods, and for methods directly based on sequence similarity. ``Distance'' employs as a pairing score the Hamming distance between a possible pair AB of the testing set and its closest neighbor in the training set. In ``Orthology'', one finds reciprocal closest neighbors in terms of Hamming distance (``orthologs'') for each possible within-species pair AB of the testing set, among the correct and incorrect within-species pairs of the training set. Specifically, a pair P (from the testing set) and a pair P' (from the training set) are considered as reciprocal closest neighbors if (i) P is the the closest neighbor of P' among the pairs of the species of P and (ii) P' is the closest neighbor of P among all the pairs from the training set. Next, one ranks possible pairs AB of the testing set using this notion of orthology: the pairs whose orthologs are all correct pairs from the training set come first, ranked by decreasing number of orthologs. Then, the pairs of the testing set that have both correct and incorrect orthologs are ranked by decreasing fraction of correct pairs among their orthologs. Finally, the pairs whose orthologs are all incorrect are ranked by increasing number of number of orthologs. In case of equality, pairs are ranked by decreasing distance to the closest correct pair from the training set. We employ the rank of a pair AB in this scheme as a pairing score. The four pairing scores corresponding to each of the four methods are employed in two ways: either within each species we find the B chain with optimal pairing score with each A chain (dashed lines), or within each species we employ the Hungarian matching algorithm to find the one-to-one pairing of A and B chains that optimizes the sum of the pairing scores (``Hung.'', solid lines). Each species comprises 4 chains AB. \textbf{B: } Fraction of pairs correctly identified (TP fraction) versus number of pairs per species, employing the same methods (and same colors) as in panel A, and a training set of 50\% of the total dataset. For the Orthology method, results are shown only up to 8 pairs per species, because computations become lengthy. In both panels, and as in Fig.~\ref{Fig3}, data was generated using a tree of 10 generations, with exactly 5 mutations per branch, out of 200 bits in each chain AB, thus yielding 1024 chains AB. Species were built randomly, and some of them were chosen randomly to build the training set, the remaining ones making up the testing set. The chance expectation, corresponding to random within-species pairings, is shown for comparison. Results are averaged over 100 replicates in panel A and 20 replicates in panel B, each corresponding to a different realization of the branching process used for data generation. }
	\label{FigS4}
\end{figure}

\begin{figure}[h!]
	\centering
	\includegraphics[width=\textwidth]{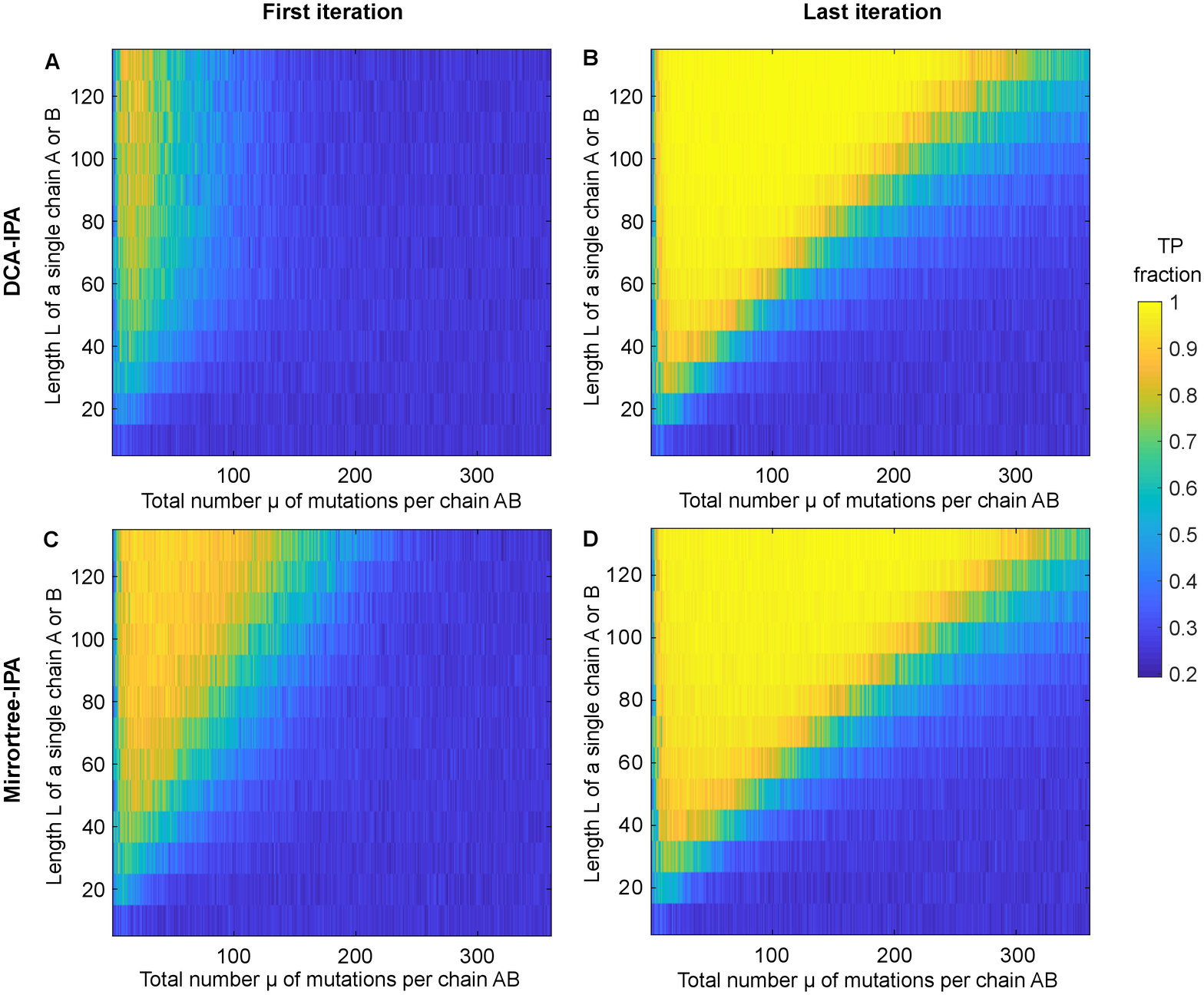}
	\vspace{0.2cm}
	\caption{{\bf Performance of the DCA-IPA and of the Mirrortree-IPA.} The fraction of pairs AB correctly predicted (TP fraction) is shown versus the total number $\mu$ of mutations per chain AB and the length $L$ of a single chain A or B for the DCA-IPA at the first (\textbf{A}) and last iteration (\textbf{B}), and similarly for the Mirrortree-IPA (\textbf{C} and \textbf{D}). Predictions were made without any training set, and the first iteration employs random within-species pairings to compute the initial pairing scores. An increment step $N_\textrm{increment}=100$ was used. The Hungarian algorithm was employed to predict pairings. Data was generated using a tree with 10 generations, with a variable average number of mutations per branch and a variable chain length, thus yielding 1024 chains AB, and random species with 4 chains AB each were constructed. }
	\label{FigS5}
\end{figure}

\begin{figure}[h!]
	\centering
	\includegraphics[width=\textwidth]{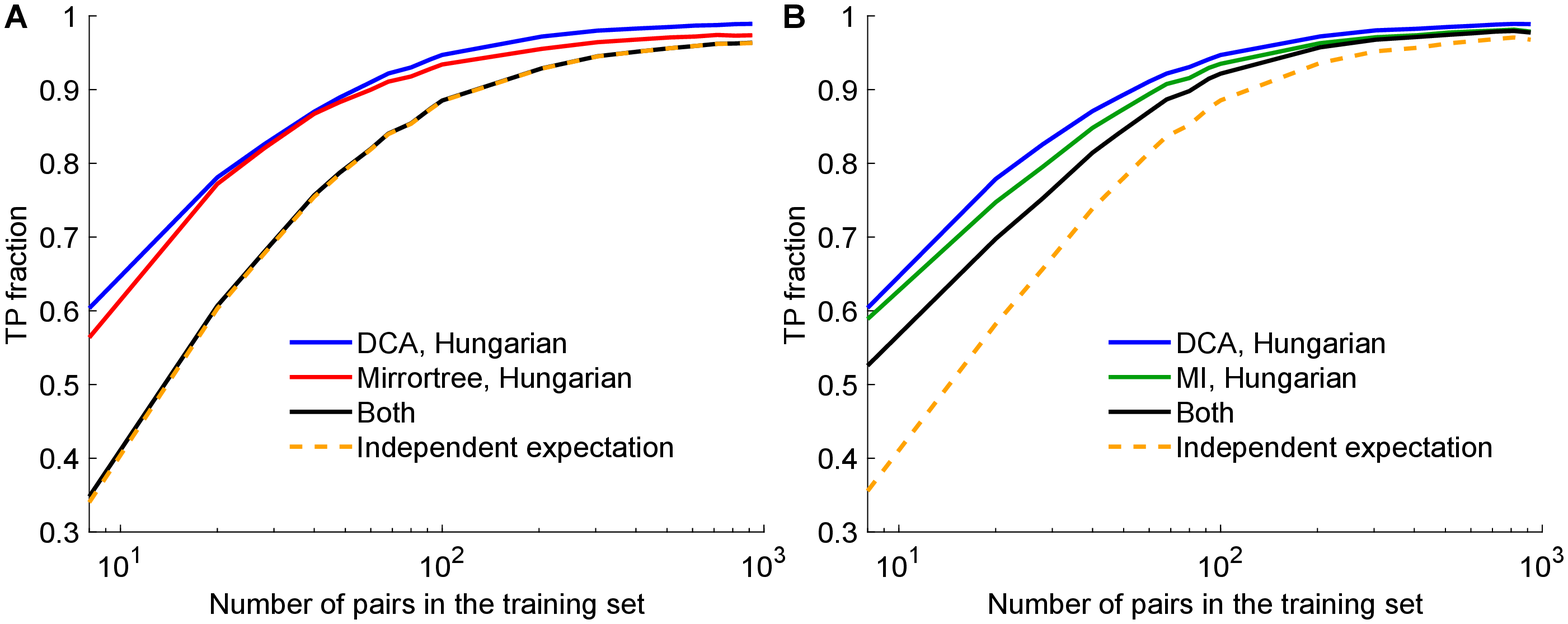}
	\vspace{0.2cm}
	\caption{{\bf Correlation of the predictions from different methods.} \textbf{A:} The fraction of pairs AB correctly predicted (TP fraction) is shown versus the number of pairs in the training set, for DCA- and Mirrortree-based pairing methods. The fraction of pairs correctly predicted by both methods is also shown, as well as its expectation if the two methods were fully independent. \textbf{B:} Similar plot, comparing DCA and MI instead of DCA and Mirrortree. In both panels, data was generated using a tree of 10 generations, with 5 mutations per branch on average, out of 200 bits in each chain AB, thus yielding 1024 chains AB. Random species with 4 chains AB each were constructed. Results are averaged over 100 replicates, each corresponding to a different realization of the branching process used for data generation.}
	\label{FigS6}
\end{figure}

\begin{figure}[h!]
	\centering
	\includegraphics[width=\textwidth]{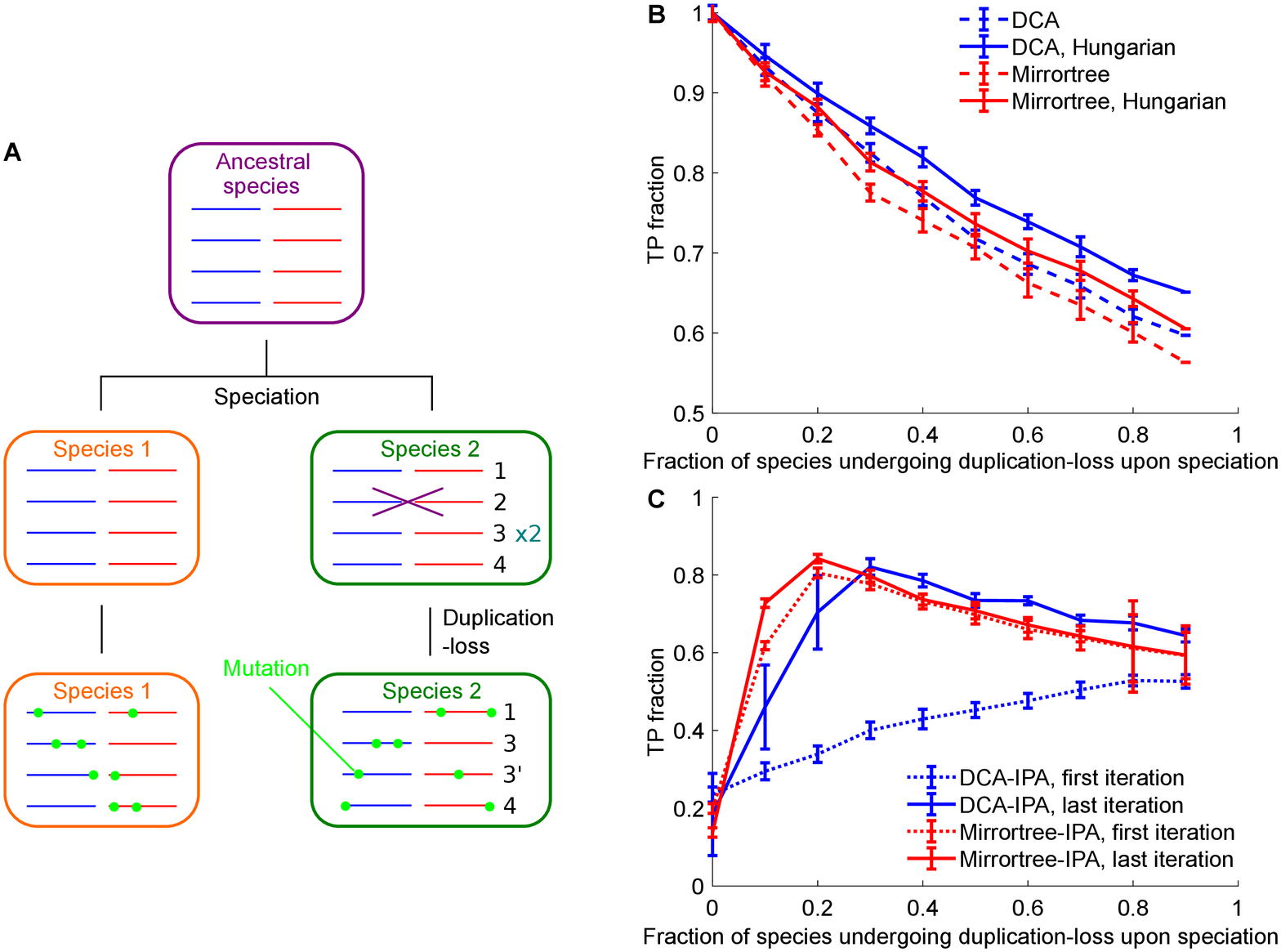}
	\vspace{0.2cm}
	\caption{{\bf Extension to different phylogenies.} \textbf{A:} Starting from one ancestral species with 4 random chains whose two halves A and B are shaded in blue and red, a series of speciation and diversification steps are performed (each sequence of these steps is called a ``generation''). Upon speciation, each species is duplicated. Then, each of the two resulting species has a certain probability to undergo a duplication-loss step where one chain is eliminated and replaced by a copy of another chain from the same species. Next, mutations occur independently in each species (light green; here 2 bits per chain AB are mutated at each generation). Hence, the number of species doubles at each generation. \textbf{B:} Predictions with a training set. The fraction of pairs AB correctly predicted (TP fraction) is shown versus the fraction of species that undergo duplication-loss at each generation, for DCA- and Mirrortree-based pairing methods. Two versions are shown for each method: either within each species we find the B chain with optimal pairing score with each A chain (dashed lines), or within each species we employ the Hungarian matching algorithm to find the one-to-one pairing of A and B chains that optimizes the sum of the pairing scores (solid lines). Here, 25 species were randomly selected to form a training set of 100 pairs AB employed to build the pairing scores, and the rest constitutes the testing set. \textbf{C:} Prediction without a training set. The fraction of pairs AB correctly predicted (TP fraction) is shown versus the fraction of species that undergo duplication-loss at each generation, for the DCA-IPA and the Mirrortree-IPA, at the first and last iterations. An increment step $N_\textrm{increment}=100$ is used. The Hungarian algorithm is employed to predict pairings. In panels B and C, data was generated using a tree of 8 generations, with 5 mutations per branch on average, out of 200 bits in each chain AB. Because here we start from one ancestral species with 4 random chains and not from 1 single random chain, this yields 1024 chains AB, already separated in species. Results are averaged over 20 replicates, each corresponding to a different realization of the branching process used for data generation. Error bars represent 95\% confidence intervals. 
		}
	\label{FigS7}
\end{figure}

\begin{table}[h!]
	
	\centering
	\begin{small}	
	\begin{tabular}{| l | l | p{2.5cm} | p{2.5cm} | p{4.5cm} |}
		\hline
		Protein 1 & Protein 2 & Mirrortree score (Pearson correlation)~\cite{Pazos01,Ochoa10} & pMirrortree score (p-value)~\cite{Ochoa15}& Comments\\ 
		\hline
		\hline
		LOLC &MACA & 0.82 & 0.76 &\\
		\hline
		ACRE &ENVR & 0.82 & 0.24 & pMirrortree score was calculated between ACRE and ACRR (close paralog of ENVR).  \\
		\hline
		\hline
		BASS (HK) &BASR (RR) & 0.96 & 0.003 &  \\
		\hline
		MALG &MALK & 0.85 & 0.84 & pMirrortree score is 0.013 when considering the close paralogs POTI-POTA of MALG-MALK.  \\
		\hline
		XDHA &XDHC & 0.86 & 0.002 &\\
		\hline
	\end{tabular}	
	\end{small}
	\vspace{0.4cm}
	\caption{\textbf{Mirrortree results for the protein pairs considered in Fig~\ref{Fig6}.} For each pair of \textit{E. coli} proteins, we applied the original Mirrortree and pMirrortree methods, as described in Refs.~\cite{Pazos01,Ochoa10} and~\cite{Ochoa15}, respectively. The resulting scores quantify the similarity of the phylogenetic trees constructed from the orthologs of the two proteins considered. Note that paralogs are not included in this analysis, in contrast to the one presented in our paper. The Mirrortree score is a Pearson correlation between the sets of distances of the two protein families considered, and a large value is indicative of coevolution~\cite{Pazos01,Ochoa10}. Among the five pairs considered, LOLC-MACA and ACRE-ENVR have the smallest Mirrortree scores, in agreement with the fact that they are the only pairs that possess no known direct physical interactions. The pMirrortree score is a p-value that provides an assessment of the confidence in the tree similarity scores: a small pMirrortree score is indicative of coevolution~\cite{Ochoa15}. Considering the standard significance threshold 0.05, BASS-BASR (HK-RR) and XDHA-XDHC have significant coevolution, while LOLC-MACA and ACRE-ENVR do not. This is in good agreement with the fact that BASS-BASR and XDHA-XDHC possess known direct physical interactions, while LOLC-MACA and ACRE-ENVR do not. Surprisingly, MALG-MALK has a large pMirrortree score while possessing a known direct physical interaction, but this is mitigated by the fact that the close paralogs POTI-POTA have significant coevolution according to their pMirrortree score.}
	\label{TableS1}
\end{table}

\clearpage


%
%
%
\bibliographystyle{plos2015}

\end{document}